\documentclass{SciPost}

\usepackage{subcaption}
\usepackage[bitstream-charter]{mathdesign}
\usepackage{bm}
\usepackage{cancel}
\newcommand{\lm}[1]{\textcolor{red}{#1}}

\usepackage[normalem]{ulem}
\usepackage{xcolor}
\usepackage{tabularx}
\usepackage{booktabs}

\renewcommand{\lm}[1]{#1}
\renewcommand{\textcolor}[2]{#2}

\DeclareSymbolFont{usualmathcal}{OMS}{cmsy}{m}{n}
\DeclareSymbolFontAlphabet{\mathcal}{usualmathcal}

\fancypagestyle{SPstyle}{
\fancyhf{}
\lhead{\colorbox{scipostblue}{\bf \color{white} ~SciPost Physics }}
\rhead{{\bf \color{scipostdeepblue} ~Submission }}

\fancyfoot[C]{\textbf{\thepage}}
}

\hypersetup{
    colorlinks=true,
    linkcolor={red!50!black},
    citecolor={blue!50!black},
    urlcolor={blue!80!black}
}

\begin{document}

\pagestyle{SPstyle}

\begin{center}{\Large \textbf{\color{scipostdeepblue}{Lindbladian approach 
 for  many-qubit  thermal machines: enhancing the performance with geometric heat pumping by \textcolor{red}{interaction}\\
}}}\end{center}

\begin{center}
\textbf{
Ger\'onimo J. Caselli\textsuperscript{1$\star$},
Luis O. Manuel\textsuperscript{1} and
Liliana Arrachea\textsuperscript{2$\dagger$}
}
\end{center}

\begin{center}
{\bf 1} Facultad de Ciencias Exactas, Ingenier\'ia y Agrimensura, Universidad Nacional de Rosario and Instituto de F\'isica Rosario (CONICET), Rosario, Argentina\\
{\bf 2} Departamento de S\'olidos Cu\'anticos y Sistemas Desordenados, Centro At\'omico Bariloche, Instituto de Nanociencia y Nanotecnolog\'ia CONICET-CNEA and Instituto Balseiro (8400), San Carlos de Bariloche, Argentina.
\\[\baselineskip]
$\star$ \href{mailto:email1}{\small caselli@ifir-conicet.gov.ar },\quad $\dagger$ \href{mailto:email2}{\small liliana.arrachea@ib.edu.ar }
\end{center}

\section*{\color{scipostdeepblue}{Abstract}}
We present a detailed analysis of slowly driven quantum thermal machines based on 
interacting qubits within the framework of the Lindblad master equation. 
By implementing a systematic expansion in the driving rate, we derive explicit expressions for the rate of work of the driving forces, the heat currents exchanged with the reservoirs, and the entropy production up to second order, 
ensuring full thermodynamic consistency in the linear-response regime. 
The formalism naturally separates geometric and dissipative contributions, identified by a Berry curvature 
and a metric in parameter space, respectively. 
Analytical results show that the geometric heat pumped per cycle is bounded by $k_B T N_q \ln 2$ for $N_q$ 
non-interacting qubits, in direct analogy with the Landauer limit for entropy change. 
This bound can be surpassed when qubit interactions and asymmetric couplings to the baths are introduced. 
Numerical results for the interacting two-qubit system reveal a non-trivial role of the interaction between qubits and the coupling between the qubits and the baths in the behavior of the dissipated power. 
The approach provides a general platform for studying dissipation, pumping, and performance optimization in driven quantum devices operating as heat engines.

\vspace{\baselineskip}

%%%%%%%%%% BLOCK: Copyright information
% This block will be filled during the proof stage, and finilized just before publication.
% It exists here only as a placeholder, and should not be modified by authors.
\noindent\textcolor{white!90!black}{%
\fbox{\parbox{0.975\linewidth}{%
\textcolor{white!40!black}{\begin{tabular}{lr}%
  \begin{minipage}{0.6\textwidth}%
    {\small Copyright attribution to authors. \newline
    This work is a submission to SciPost Physics. \newline
    License information to appear upon publication. \newline
    Publication information to appear upon publication.}
  \end{minipage} & \begin{minipage}{0.4\textwidth}
    {\small Received Date \newline Accepted Date \newline Published Date}%
  \end{minipage}
\end{tabular}}
}}
}
%\linenumbers


\vspace{10pt}
\noindent\rule{\textwidth}{1pt}
\tableofcontents
\noindent\rule{\textwidth}{1pt}
\vspace{10pt}

\section{Introduction}
\label{sec:intro}

The study of small-scale thermal machines, energy dissipation and conversion in these systems, has become 
an active avenue of research for some time now. 
The area is identified as {\em quantum thermodynamics} \cite{Vinjanampathy2016Oct,Goold2016Feb,binder2018thermodynamics,Mitchison2019Apr,Campbell2025Nov}, 
where stochastic statistical physics \cite{Jarzynski2011Mar,Seifert2012Nov}, quantum information, photonic, atomic 
and molecular physics, and condensed-matter physics converge. The goal is to understand heat-work conversion~\cite{Benenti2017Jun,Myers2022Apr,Pekola2021Oct,Arrachea2023Jan,Cangemi2024Oct}, as well as entropy-production\cite{Landi2021Sep} and fluctuation mechanisms\cite{Manzano2022Jun}.

Thermodynamic cycles are paradigmatic processes to implement heat-work conversion.  Several proposals and experiments in the quantum realm are based on a few-level system, like an atom or a qubit, operating in   strokes  \cite{vonLindenfels2019Aug,Rossnagel2016Apr,Ronzani2018Oct,Rezakhani2009Aug,Schmiedl2007Dec,Kosloff2017Mar,Palao2001Oct}, as in the case of classical textbook examples. In these protocols, 
the system couples to thermal baths sequentially, typically to a hot and a cold one.
 An important aspect in the optimization of the performance of these machines is to take into account the
 finite time of the operation at each stroke. This implies considering the entropy production and  the dissipation, as well as  the power at which useful work is generated in a heat engine, or heat is extracted in a refrigerator. Such
 effects have been addressed in Refs. \cite{Abiuso2020Mar,Brandner2020Jan,Miller2020Dec,Miller2021May} within the slow-driving regime, characterized by operation time scales much longer than the typical time scale of the
internal dynamics of the quantum system when in equilibrium with the baths. In this regime,
an insightful perspective consists in describing  the dissipated power in terms of a metric in the space of the driving parameters. This enables the definition  of the thermodynamic length and the definitions of bound for the dissipation, as 
originally introduced in the context of classical systems \cite{Weinhold1975Sep,Salamon1980Jul,Salamon1983Sep,Crooks2007Sep} and later generalized to quantum settings \cite{Scandi2019Oct}.

 Quantum pumping is another cyclic process of a few-level system in contact to two or more reservoirs of particles 
 or energy. It consists of a periodic change of two or more parameters of the system, in combination with some asymmetry 
 in the setup or in the driving protocol, resulting in a net transfer of particles and/or energy between the reservoirs, 
 even in the absence of a chemical potential or thermal bias.
 This phenomenon has been widely studied in the context of electron systems, both experimentally \cite{Pothier1992Jan,Switkes1999Mar} and theoretically\cite{Brouwer1998Oct,Altshuler1999Mar,
Moskalets2004May,Splettstoesser2006Aug,Arrachea2006Dec}, and it is central to the 
study of energy pumping in spin and nanomechanical systems 
(see Ref. \cite{Arrachea2023Jan,Acciai2025Oct} for an overview). 
A relevant regime corresponds to slow driving, where  geometric properties play a significant role in the characterization of this mechanism \cite{Brouwer1998Oct,Avron2000Oct,Avron2004Aug,Moskalets2004May,Splettstoesser2006Aug,Arrachea2006Dec,Ren2010Apr,Wang2022Feb,Esin2025Apr}. 
The net pumped charge or heat per cycle is described by the integral of a generalized Berry curvature \cite{Victor1984Mar}. 
 
 In combination with a chemical potential or temperature bias imposed at the reservoirs, 
 the cyclic time-dependent protocol defines, respectively, a motor  or a thermal machine. 
 For slow driving and small bias, it can be shown that \cite{Bustos-Marun2013Aug,
 Ludovico2016Feb,Bhandari2020Oct}
 quantum pumping provides the mechanism of heat-work conversion characterized by a geometric quantity. These machines  
 operate in permanent contact with the reservoirs and  pumping takes place 
 simultaneously with the heat and/or particle leak and the dissipation, as in thermoelectricity \cite{Benenti2017Jun}. 
 
 The role of geometric properties characterizing both pumping and dissipation in quantum thermal machines was analyzed in Ref. \cite{TerrenAlonso2022Feb}.
 It was shown that the maximal generated power 
 in a heat engine operating under slow driving by two parameters, corresponds to the
 optimal ratio between an effective area weighted by the Berry curvature associated with
 pumping and the thermodynamic length defined  by the metric associated with
 dissipation. The optimization of specific protocols for a qubit system 
 driven by two parameters was also addressed in that work on the basis of a master quantum equation derived in Ref. \cite{Bhandari2021Jul}.
 
 The aim of the present contribution is twofold: first, we present a treatment to analyze pumping and 
 dissipation in quantum thermal machines under slow driving based on the Lindblad master equation; 
 second, we analyze the case of $N_q$ interacting qubits and identify the importance of \lm{correlation} in enhancing energy pumping between reservoirs.  We show that, for a single driven qubit, 
 the results coincide with those of Ref. \cite{TerrenAlonso2022Feb}. 
 In addition, we show that, for non-interacting qubits, the heat pumping is bounded by $k_B T N_q \ln 2$, 
 which is $N_q$ times the Landauer bound obtained in Ref. \cite{TerrenAlonso2022Feb} for a single qubit. 
 However, this bound is no longer valid for interacting qubits and can be exceeded by suitable protocols and appropriate couplings to the thermal baths.

The paper is organized as follows. 
In Section 2 we present the slow-driving expansion to describe the dynamics of the many-qubit driven system.
We present the model,  the expressions for the power and the heat current, we analyze the thermodynamic consistency, and we introduce 
the geometric properties in the operation of slowly-driven thermal machines. 
Analytical results for the bound on the pumped heat between reservoirs are also presented, together with its connection to the Landauer bound for the entropy change \cite{Landauer1961Jul,Landauer1988Oct}.
Section 3 summarizes the description of the driven system coupled to reservoirs at different temperatures within the linear response description \cite{Bhandari2020Oct}. Section 4 is devoted to numerical results for the particular case of two interacting qubits coupled to two reservoirs. We present benchmarks for the equations describing the energy and entropy balance. We also present results for the heat pumping, dissipation,  and the figure of merit that qualifies the performance of the machine as a heat engine. We analyze in detail the role of the interaction between qubits and the couplings between qubits and reservoirs. 
Section 5 is devoted to discussion and conclusions. Finally, a series of appendices are included at the end of the paper, where the most relevant derivations and analytical developments are provided.

\section{Lindbladian approach for many-qubit thermal machine}\label{theoretical_framework}
\textcolor{red}{The derivation of completely positive Markovian quantum master equations  to describe time-dependent problems from first principles has been the subject of many previous contributions. Within the slow-driving regime, a framework was introduced in Ref. \cite{Albash2012Dec}, followed by a Kubo-type linear response approach developed in Ref. \cite{CamposVenuti2016Mar}. For strongly non-equilibrium conditions, alternative schemes were presented in Refs. \cite{Dann2018Nov,Mozgunov2020Feb}.}

In our case, we focus on the slow-driving regime and Hamiltonians for the few-level system that depend on on time though a set of control parameters. We follow Ref. \cite{Cavina2017Aug}, which relies on a perturbative expansion around the quasistatic stationary solution of the Lindblad master equation. This approach assumes that the  control parameters vary on a time scale much longer than the typical time scale governing the internal dynamics of the system coupled to the baths. We first introduce the model for the system of driven qubits coupled to reservoirs, the Lindblad master equation for the problem with frozen parameters,

\subsection{Model}

We consider the following Hamiltonian for  $N_{\rm q}$ coupled qubits under slow driving
\begin{equation}\label{ham_sys}
{\cal H}_{\rm S}(t) = -\sum_{j=1}^{N_{\rm q}} \boldsymbol{B}_j(t) \cdot {\bf S}_j + 
J \sum_{i=1}^{N_{\rm q}-1}{\bf S}_j \cdot {\bf S}_{j+1},
\end{equation}
where the qubits are modeled in terms of spin 1/2 operators  ${\bf S}_j$, which are individually operated by  time-dependent parameters $\boldsymbol{B}_j(t)$, while  $J$ is the exchange interaction between them.

This system is coupled to $M$ thermal baths, each described by a bosonic Hamiltonian,
% \begin{equation}
% {\cal H}_{\alpha} = \sum_{k_\alpha} \omega_{k_\alpha} b^{\dagger}_{k_\alpha} b_{k_\alpha}, \;\;\;\;\;\; \alpha=1,\ldots, M.
% \end{equation}

\begin{equation}
{\cal H}_{\alpha} = \sum_{k} \omega_{k \alpha} b^{\dagger}_{k \alpha} b_{k \alpha}, \;\;\;\;\;\; \alpha=1,\ldots, M,
\end{equation}
where $\omega_{k\alpha}$ denote the energies of the normal modes of bath~$\alpha$, and 
$b_{k\alpha}$ ($b^{\dagger}_{k\alpha}$) annihilates (creates) an excitation in mode~$k$.

The Hamiltonians describing the coupling between the system and the thermal baths read
\begin{equation}\label{c-bath}
{\cal V}_{\alpha} = g_\alpha \pi_\alpha B_\alpha, \qquad \alpha=1,\ldots, M,
\end{equation}
where $\pi_\alpha$ is a Hermitian operator of the system, and $B_\alpha$ is a Hermitian bath operator defined as a linear combination of the normal modes of bath~$\alpha$, 
\begin{equation}
B_\alpha = \sum_{k} \left(r_{k \alpha} b_{k \alpha} + r^{*}_{k\alpha} b^\dagger_{k\alpha}\right),
\label{eq:erres}
\end{equation}
and $g_{\alpha}$ sets the overall system–bath coupling strength.

\subsection{Lindblad quantum master equation for the frozen system}

We focus on the weak-coupling limit between the driven system and the reservoirs, where the 
time-dependent parameters are frozen at an observational time $t$, specified by the external driving field $\bm B(t)$. 
In Appendix~\ref{Lindblad_derivation}, we show that performing the usual approximations, 
the reduced density matrix of the system obeys the following Lindblad master equation,
\begin{equation} \label{lind}
    \frac{d\rho^{(f)}(t;s)}{ds}
    = -i \big[\mathcal{H}_{\mathrm{S}}(t),\,\rho^{(f)}(t;s)\big]
    + \sum_{\alpha} \mathcal{D}_{\alpha}\!\left[\rho^{(f)}(t;s)\right],
\end{equation}
where the auxiliary time variable $s$ describes the non-unitary evolution, ruled by the dissipators 
associated with the frozen system Hamiltonian ${\cal H}_{\rm S}(t)$. 
% \lmc{Agregaría algo como que tiene Cavina, estamos considerando que la ecuación de arriba tiene una única solución estacionaria, que dicha solución es la que estamos luego expandiendo en el tiempo $t$. }
The corresponding dissipators read
\begin{equation}
{\cal D}_{\alpha}\!\left[\rho^{(f)}\right] =
\sum_{\omega}    L_{\alpha\omega}\, \rho^{(f)}\, L^{\dagger}_{\alpha\omega}
    - \frac{1}{2} \Big\{
        L^{\dagger}_{\alpha\omega}L_{\alpha\omega},\,
        \rho^{(f)}
      \Big\}.
\end{equation}

% \begin{equation}
%     \mathcal{D}_{\alpha}[\rho^{(f)}] = \sum_{\lambda=\pm} \sum_\omega L_{\alpha \lambda}(\omega) \; \rho^{(f)} \; L^\dagger_{\alpha \lambda}(\omega) - \frac{1}{2} \left( L^\dagger_{\alpha \lambda}(\omega) L_{\alpha \lambda}(\omega) \: \rho^{(f)} + \rho^{(f)} \: L^\dagger_{\alpha \lambda}(\omega) L_{\alpha \lambda}(\omega) \right).
% \end{equation}

The jump operators $L_{\alpha\omega}$ are defined with respect to the instantaneous eigenbasis 
of the system  Hamiltonian, ${\cal H}_{\rm S}(t)\,|m\rangle = \epsilon_m |m\rangle$. They can be written as
\begin{equation}
    L_{\alpha \omega} = g_\alpha \sqrt{\gamma_\alpha(\omega)}\, \pi_{\alpha \omega}, \;\;\;\;\;\;\;\;\:\: \pi_{\alpha \omega}= \sum_{\substack{l,m \\ \epsilon_{ml} = \omega\textcolor{red}{>0}}}
      \xi^{\alpha}_{lm}\,|l\rangle\langle m|,
      \label{L-ops}
\end{equation}
where $\pi_{\alpha \omega}$ annihilates (for $\omega>0$) or creates (for $\omega<0$) an energy quantum $|\omega|$ in the system through its coupling to reservoir $\alpha$. 
Here, $\epsilon_{lm} = \epsilon_l - \epsilon_m$, and 
$\xi^{\alpha}_{lm} = \langle l|\pi_\alpha|m\rangle$ are the matrix elements of the coupling operator $\pi_\alpha$ in the instantaneous eigenbasis. 
The transition rate function $\gamma_\alpha(\omega)$ is the real part of the bath correlation function associated with 
the operator $B_\alpha$.
For Ohmic baths, and for an appropriate choice of the coupling coefficients $r_{\alpha k}$ in Eq.~\eqref{eq:erres}, 
\begin{equation}
\gamma_{\alpha}(\omega) = 
\begin{cases}
 \omega\,(1+n_\alpha(\omega))\, e^{-\omega/\omega_C}, & \omega > 0, \\[4pt]
|\omega|\, n_{\alpha}(|\omega|)\, e^{-|\omega|/\omega_C}, & \omega < 0,
\label{gammas}
\end{cases}
\end{equation}
where $n_\alpha(\omega) = [e^{\beta_\alpha \omega} - 1]^{-1}$ is the Bose distribution,
$T_\alpha = 1/(k_B \beta_\alpha)$ is  the reservoir's temperature, and $\omega_C$ denotes the high-frequency cutoff of the bath spectrum.

% \begin{eqnarray} \label{L-ops}
%     L_{\alpha +}(\omega) &=& \sum_{\substack{l,m \\ \epsilon_{lm}=\omega}} \xi^\alpha_{lm} \sqrt{\Gamma_\alpha(\epsilon_{lm})} |l\rangle \langle {m}|, \nonumber \\
%     L_{\alpha -}(\omega)  &=& \sum_{\substack{l,m \\ \epsilon_{lm}=\omega}} \xi^\alpha_{lm} \sqrt{\overline{\Gamma}_\alpha(\epsilon_{ml})} |l\rangle \langle {m}|,
% \end{eqnarray}
% where  $\epsilon_{lm}=\epsilon_l-\epsilon_m$. \lm{$\xi^{\alpha}_{lm}$ are the matrix elements of $\pi_\alpha$ between the instantaneous eigenstates: $\xi^{\alpha}_{lm} = \langle l|\pi_\alpha|m\rangle$.} 
% % The operators entering the contact
% %  are transformed to this basis by means of a unitary transformation $U$ as follows,
% % $\xi_\alpha = U^\dagger\; \pi_\alpha \;U$. 
% %
% The transition rates are
% \begin{eqnarray}\label{gammas}
%     \Gamma_\alpha(\omega) &=& \gamma_\alpha(\omega)\; n_\alpha(\omega),\nonumber \\
%     \overline{\Gamma}_\alpha(\omega) &=& \gamma_\alpha(\omega) \;(n_\alpha(\omega)+1),
% \end{eqnarray}
% where $n_\alpha(\omega)=[e^{\beta_\alpha\omega}-1]^{-1}$ is the Bose-Einstein distribution function, which depends on the temperature $T_\alpha=1/k_B \beta_\alpha$ of the corresponding reservoir. We consider
% Ohmic density of states with a cutoff $\epsilon_C$ for these systems, which defines the functions 
% $\gamma_\alpha(\omega) = g_\alpha \; \Theta(\omega) \; \omega \; e^{-\omega/\omega_C}$. 

Importantly, the rates are evaluated at second order in the coupling constants between the system and  the reservoirs and there are no
processes involving simultaneously two different reservoirs. 
For this reason, the dissipator entering Eq. (\ref{lind}) is a linear combination of the dissipators associated with 
the different baths.

\subsection{Slow-driving expansion}
We consider all baths at the same temperature, $T_\alpha = T$. This assumption allows us to isolate and analyze the 
effects that arise solely from the time-dependent driving, without introducing additional energy fluxes associated 
with temperature differences.

We identify the control parameters with a vector $\bm X(t)$.
In the case of the Hamiltonian Eq.~\eqref{ham_sys}, each time-dependent component of the vectors
$\bm B_j(t)$ defines a component of this vector, $\bm X(t)=\left(B_{1,x}(t),\ldots, B_{N_{\rm q},z}(t)\right)$.

Following Refs.~\cite{Cavina2017Aug,Scandi2019Oct}, \textcolor{red}{we  assume that for all values of the control parameters, there exist a unique stationary solution $\rho^{(f)}(\bm X)$
of
Eq. (\ref{lind}), associated with the frozen  Hamiltonian ${\cal H}_{\rm S}(\bm X)$, where the parameters $\bm X$ are evaluated at a fixed time, 
which satisfies
\begin{equation}
  \frac{d\rho^{(f)}(\bm X;{s }) }{d{s }}= \sum_\alpha{\cal D}_\alpha \; \left[\rho^{(f)}(\bm X ;{s})\right] = 0.
  \label{eq:rhof1}
\end{equation}
Next, we consider slow changes of these parameters, assuming that they are characterized by a time scale $\tau$ --the period in the case of periodic driving-- that is much longer than the characteristic time scale  of the system-reservoir dynamics. In the case of periodic driving, this is equivalent to consider driving frequencies satisfying that $\hbar /\tau$ is much lower than any energy scale of the few-level system coupled to the reservoir.}
We introduce an expansion in powers of ${\tau}^{-1}$ of the reduced density matrix describing the dynamics of the time-dependent system coupled to the baths,
\begin{equation}\label{slow-rho}
    \rho(t)=\rho^{(f)}(\bm X)+\rho^{(1)}(\dot{\bm X})+ \rho^{(2)}(\dot{\bm X},\ddot{\bm X})+\ldots.
\end{equation}
Here, $\rho^{(f)}(\bm X)$ denotes the reduced density operator associated with the frozen  Hamiltonian ${\cal H}_{\rm S}(\bm X)$, where the parameters $\bm X$ are evaluated at a fixed time. The other terms represent corrections proportional to the "velocities" $\dot{\bm X}$ and "accelerations" $\ddot{\bm X}$. 
Although this expansion can be extended to higher orders, 
we focus here on contributions up to  order ${\tau}^{-2}$. 
The first-order term defines the so-called adiabatic response approximation.

%The first term is the stationary solution of Eq. (\ref{lind}), which reduces to
%
%\begin{equation}
%  \frac{d\rho^{(f)}(\bm X;{s }) }{d{s }}= \sum_\alpha{\cal D}_\alpha \; \left[\rho^{(f)}(\bm X ;{s})\right] = 0.
%  \label{eq:rhof1}
%\end{equation}
%
We decompose the frozen density operator as follows
\begin{equation}
    {\rho}^{(f)}(\bm X) =\frac{1}{N}\mathbb{I}+\bar{\rho}^f(\bm X),
\end{equation}
where $\mathrm{Tr}\left[\bar{\rho}^{(f)}(\bm X)\right]=0$, and $N$ denotes the number of energy levels of the system.
We also introduce the  Lindbladian  operator acting on  the traceless subspace, $\mathrm{Tr}\;\bar{\rho} =0$,
\begin{equation}
 {\cal L}_f  \left[{\bar \rho}\right]    \equiv \sum_{\alpha}{\cal D}_\alpha \left[\bar{\rho}\right].
 %{\cal L}_f  {\tilde \rho}^{(f)}(\bm X)    \equiv \sum_{\alpha}{\cal D}_\alpha \tilde{\rho}^{(f)}(\bm X).
\end{equation}

 With these definitions, the other terms of the expansion Eq. (\ref{slow-rho}) are calculated from 
 %\textcolor{blue}{Le puse de nuevo el tilde a la parte traceless de $\rho^f$, yo se lo habia quitado por error. Está bien asi, trabajamos con la parte traceless.}
\begin{align}\label{slow-exp}
\rho^{(1)}(\bm X)&=  \left[{\cal L}_f^{-1} \; \frac{d}{dt}\right]  {\rho}^{(f)}(\bm X) = \left[{\cal L}_f^{-1} \; \partial_{\bm X}\right]{\rho}^{(f)}(\bm X) \cdot \dot{\bm X},\nonumber \\
%{\cal L}_f^{-1} \;  U(\bm X)\left\{\partial_{\bm X} \left[ U^\dagger(\bm X) \tilde{\rho}^{(f)}(\bm X) U(\bm X)\right] \cdot \dot{\bm X}U^\dagger(\bm X)\right\} ,\nonumber \\
\rho^{(2)}(\bm X) &= \left[{\cal L}_f^{-1} \; \frac{d}{dt} \right] \rho^{(1)}(\bm X) \cdot \dot{\bm X} \nonumber \\
%&= \left[{\cal L}_f^{-1} \; \partial_{\bm X}\right]^2 \tilde{\rho}^{(f)}(\bm X) \: \dot{\bm X}\cdot\dot{\bm X} +  \left[{\cal L}_f^{-1} \; \partial^2_{\bm X}\right]  \tilde{\rho}^{(f)}(\bm X) \cdot \ddot{\bm X} \\
%&= {\cal L}_f^{-1} \; \left( \frac{d}{dt} \rho^{(1)}(\bm X) \right)  \cdot \dot{\bm X} + \left[{\cal L}_f^{-1} \right] \rho^{(1)}(\bm X) \cdot  \ddot{\bm X} \\
&= \dot{\bm X} \cdot \left[{\cal L}_f^{-1} \; \partial_{\bm X}\right]^2  \rho^{(f)}(\bm X) \cdot \dot{\bm X} + \left[{\cal L}_f^{-1}\right]^2 
%{\cal L}_f^{-1} 
\; \partial_{\bm X} \rho^{(f)}(\bm X)  \cdot  \ddot{\bm X}.
\end{align}
We notice that these components satisfy order by order the following Lindblad equation 
\begin{equation}\label{lindt}
\frac{d \rho^{(n-1)}(\bm X)}{dt} = {\cal L}_f \left[\rho^{(n)}(\bm X)\right],
\end{equation}
with $\rho^{(0)}(\bm X) \equiv \rho^{(f)}(\bm X) $.
%These equations follow the notation in reference \cite{Cavina2017Aug}, where $\mathcal{P}$ is the proyector into the traceless subspace of density operators, so that $(\mathcal{L}_f\mathcal{P})$ is invertible. $\mathbb{I}_d$ is the identity matrix for a Hilbert space of dimension $d$. 
%
%
Since $\rho^{(f)}(\bm X)$ is defined in terms of the instantaneous eigenstates of 
${\cal H}_S(\bm X)$, it is important to account for the $\bm X$-dependence of this basis when performing the operation 
$\partial_{\bm X}$. A convenient way to compute this operation is to transform the density operator 
 to a ''laboratory frame'' $\rho^{(f)}|_{\mathrm{lab}}=U^\dagger(\bm X)\;{\rho}^f(\bm X)\;U(\bm X)$ 
 --where the basis does not depend on $\bm X$--, take the derivative of $\rho^{(f)}|_{\mathrm{lab}}$ and 
 then move back to the eigenstate frame. 
 Hence, 
\begin{equation}
\partial_{\bm X}  {\rho}^{(f)}(\bm X) 
\equiv U(\bm X)\left\{\frac{\partial\left[ U^\dagger(\bm X) {\rho}^{(f)}(\bm X) U(\bm X)\right]}{\partial\bm X}\right\}\;  U^\dagger(\bm X).
\end{equation}

\section{\textcolor{red}{Thermodynamic properties}}
%
%
%\vspace{10mm}
%\begin{table}[!htb]
%\centering
%\renewcommand{\arraystretch}{1.25}

%\begin{tabular}{c}
%\toprule
%{\Large
%$\bm{\Lambda}_\alpha^{(1)}
%\qquad
%\underline{\bm{\Omega}}_{\alpha}^{(2)}$
%} \\[6pt]
%\midrule
%Boldface indicates vector.\\
%The underline indicates matrix.\\
%$\alpha$ labels indicate the thermal reservoir ($L,R$) associated to the magnitude. \\
%The upper index $(n)$ indicates the derivative order or velocity power of the magnitude.\\
%\bottomrule
%\end{tabular}
%\caption{Thermodynamic magnitudes. \lm{Esta tabla explica con dos ejemplos qué es cada indice}}
%\end{table}

\subsection{Power and heat current}

To establish the thermodynamic description of the driven quantum system, 
we identify the energy fluxes associated with the driving and with the 
coupling to the baths. The external sources perform power on the 
system, while the exchange of energy with each bath defines the corresponding heat currents. 
These quantities provide the basis for formulating the first law of thermodynamics 
in the slow-driving regime. The power developed by the driving sources reads
\begin{equation}\label{power}
    P(t) = \text{Tr}\bigg\{\frac{d{\cal H}_{\rm S}(t)}{dt} \rho(t)\bigg\} 
    =  \text{Tr}\bigg\{ \partial_{\bm X}{\cal H}_{\rm S}(t) \rho(t) \bigg\} \cdot \dot{\bm X},
\end{equation}
while the heat current exiting each reservoir can be calculated from 
\begin{equation}
    J_\alpha(t)= -\frac{d \langle {\cal H}_\alpha\rangle}{dt}.
\end{equation}
The minus sign indicates that, according to our convention, a positive heat current corresponds to 
energy flowing into the system.
Substituting the slow-driving expansion defined in Eq.~(\ref{slow-rho}) and keeping terms up to second 
order in ${\tau}^{-1}$, we get two distinct contributions to the power performed by the external driving 
fields on the system,
\begin{equation}\label{expansionp}
P(t) = P^{(1)}(t)+ P^{(2)}(t),
\end{equation}
being
\begin{eqnarray}\label{definitionsp}
    P^{(1)}(t) &=& \text{Tr}\bigg\{\rho^{(f)}(t) \frac{d{\cal H}_{\rm S}(t)}{dt}\bigg\}  
    =  \text{Tr}\bigg\{\rho^{(f)}(\bm X) \, \partial_{\bm X}{\cal H}_{\rm S}(\bm X)\bigg\} \cdot \dot{\bm X}, \nonumber \\
    P^{(2)}(t) &=& \text{Tr}\bigg\{\rho^{(1)}(t) \frac{d{\cal H}_{\rm S}(t)}{dt}\bigg\}
    = \dot{\bm X}^T \cdot \underline{\bm \Lambda}\!\left(\bm X\right) \cdot \dot{\bm X},
\end{eqnarray}
where $\underline{\bm \Lambda}\left( \bm X\right)$ is a matrix with elements 
\begin{equation}\label{lambda_mat}
\Lambda_{jl}\left( \bm X\right)=\mbox{Tr}\left\{\left[{\cal L}_f^{-1} \; \partial_{X_j}\right]  \rho^{(f)}(\bm X) \partial_{X_l}{\cal H}_{\rm S}(\bm X)\right\}.
\end{equation}

The first term corresponds to the conservative power associated with the quasi-static 
evolution of the system, where the system state instantaneously follows the control parameters.  
In contrast, $P^{(2)}(t)$ accounts for the non-equilibrium corrections 
generated by finite driving speeds, which give rise to dissipation and therefore constitute 
the non-conservative component of the power. 

To analyze the thermodynamic consistency of the slow-driving expansion,  we next study 
the heat currents exchanged between the system and the baths. In analogy with the power, 
the total heat current can be systematically expanded in powers of the driving rate 
${\tau}^{-1}$. 
Following the same procedure as that leading to Eq.~(\ref{lind}), we show in
Appendix~\ref{appendix_heat_current} that the expression of the heat flux 
corresponding to the frozen Hamiltonian reads
\begin{equation}\label{heatalf}
J^{(f)}_\alpha(t)=\text{Tr}\bigg\{{\cal D}_\alpha[\rho^{(f)}(t)] {\cal H}_{\rm S}(\bm X)\bigg\}.
\end{equation}
This term represents the stationary energy current flowing from reservoir $\alpha$
when the system is instantaneously in the frozen state $\rho^{(f)}(\bm X)$.
In thermal equilibrium, when all reservoirs have the same temperature,
the detailed balance condition --encoded in Eq.~(\ref{eq:rhof1})-- 
ensures that $J^{(f)}_\alpha=0$, as no net energy exchange occurs.
To capture the effects of slow driving, we extend Eq.~(\ref{heatalf}) to include the higher-order
corrections of the density matrix in Eq.~(\ref{slow-rho}).
Keeping terms up to second order in ${\tau}^{-1}$, we obtain
\begin{equation}\label{expansionj}
J_\alpha(t) = J^{(f)}_\alpha(t)+ J^{(1)}_\alpha(t)+J^{(2)}_\alpha(t),
\end{equation}
with
\begin{eqnarray}\label{definitions}
J^{(1)}_\alpha(t) &=& \text{Tr}\,\bigg\{\mathcal{D}_\alpha[\rho^{(1)}(t)] \mathcal{H}_{\rm S}(t)\bigg\} =  \bm \Lambda^{(1)}_\alpha(\bm X) \cdot \dot{\bm X}, \nonumber \\
J^{(2)}_\alpha(t) &=& \mathrm{Tr}\, \bigg\{\mathcal{D}_\alpha[\rho^{(2)}(t)] \mathcal{H}_{\rm S}(t)\bigg\} = \dot{\bm X}^T \cdot \underline{\bm \Omega}^{\lm{(2)}}_\alpha(\bm X)  \cdot \dot{\bm X} + \bm  \Lambda^{(2)}_\alpha(\bm X) \cdot \ddot{\bm X}, \label{Q2}
\end{eqnarray}
where $\bm \Lambda^{(1)}_\alpha(\bm X)$ and $\bm \Lambda^{(2)}_\alpha(\bm X)$
are vectors with components
\begin{equation}\label{lambda-al}
\Lambda^{(n)}_{\alpha,j}(\bm X)=\text{Tr}\;\bigg\{{\cal D}_\alpha \left[ \left({\cal L}_f^{-1}\right)^n  \; \partial_{X_j} \rho^{(f)}(\bm X)\right] {\cal H}_{\rm S}(\bm X)\bigg\},\quad n=1,2,
\end{equation}
while $\underline{\bm \Omega}^{(1)}_\alpha(\bm X)$ is a matrix with elements
\begin{equation}
\Omega^{\lm{(2)}}_{\alpha,j l}(\bm X) = \text{Tr}\;\bigg\{ \mathcal{D}_\alpha\left[ ({\cal L}_f^{-1} \; \partial_{X_j}) \; ({\cal L}_f^{-1} \; \partial_{X_l}) \; \rho^{(f)}(\bm X) \right] \mathcal{H}_{\rm S} \bigg\}.
\end{equation}
The first-order term, $J^{(1)}_\alpha(t)$, describes reversible response of the system to slow parameter 
variations.
Since it is linear in the driving velocity, it changes sign under time reversal. In contrast, the second-order term, $J^{(2)}_\alpha(t)$, captures the dissipative component of the 
heat current. 
Both the contribution quadratic in the driving velocity and the one proportional to the acceleration 
belong to this term, as they are second order in the driving rate and account for non-adiabatic 
corrections beyond the reversible response. \lm{Table \ref{tabla} summarizes the  coefficients defined in this section and used in what follows.}

\begin{table}[!htb]
\centering
\begin{tabularx}{\columnwidth}{lX}
\toprule
\textbf{Magnitude} & \textbf{Description} \\
\midrule
%$\bm{X}, \dot{\bm{X}}, \ddot{\bm{X}}$ & vectors of driving parameters, velocity and acceleration.\\
$\bm{\Lambda}_\alpha^{(1)}(\bm{X})$ & vector in the first order heat current $J^{(1)}_\alpha$ multiplying the velocity $\dot{\bm{X}}$.\\
$\bm{\Lambda}_\alpha^{(2)}(\bm{X})$ & vector in the second order heat current $J^{(2)}_\alpha$ multiplying the acceleration $\ddot{\bm{X}}$. \\
$\underline{\bm{\Omega}}^{(2)}_\alpha(\bm{X})$ & matrix in $J^{(2)}_\alpha$ multiplying the term quadratic in velocities $\dot{\bm{X}}^2$. \\
$\tilde{\underline{\bm{\Omega}}}^{(2)}_\alpha(\bm{X})$ & matrix in $J^{(2)}_\alpha$, arises when rewritting the acceleration term for closed trajectories. \\
$\underline{\bm{\Omega}}_\alpha(\bm{X})$ & sum $\underline{\bm{\Omega}}^{(2)}_\alpha+\tilde{\underline{\bm{\Omega}}}^{(2)}_\alpha$. \\
$\underline{\bm{\Lambda}}(\bm{X})$ & matrix in the second order power $P^{(2)}$. \\
\bottomrule
\end{tabularx}
\caption{\lm{Table of tensor magnitudes associated to heat current and power. Boldface denotes vectors, underlining indicates matrices, $\alpha=L,R$ labels the associated reservoir, and the superscript $(n)$ indicates the associated order in $(1/\tau)^n$.}}
\label{tabla}
\end{table}

\subsection{Thermodynamic consistency}
In this section, we verify that the slow-driving expansion introduced above preserves the fundamental 
thermodynamic relations. We start by examining the first law of thermodynamics, which connects the variations 
of the system's internal energy with the power and heat fluxes exchanged with the reservoirs.

\subsubsection{First law}
We start by analyzing the variation of the internal energy of the system
\begin{equation} \label{1st}
    \frac{dE_{\rm S}}{dt}= \text{Tr}\bigg\{\rho(t) \frac{d{\cal H}_{\rm S}(t)}{dt}  \bigg\}+    \text{Tr}\bigg\{\frac{d{\rho}(t)}{dt}{\cal H}_{\rm S}(t)  \bigg\}.
\end{equation}
The first term is the definition of the power given by Eq. (\ref{power}). Using the fact that the different 
terms of the expansion Eq. (\ref{slow-rho}) satisfy the Lindblad equation (\ref{lindt}), and using the definition of
the heat flux given by Eqs. (\ref{expansionj}) and (\ref{definitions}), we  write
\begin{equation}
     \text{Tr}\;\bigg\{\frac{d{\rho}(t)}{dt} {\cal H}_{\rm S}(t)  \bigg\} =
    \sum_{\alpha} \text{Tr}\;\bigg\{ {\cal D}_\alpha[\rho(t)] {\cal H}_{\rm S}(t) \bigg\}
= \sum_\alpha J_\alpha(t).
\end{equation}
Hence, Eq. (\ref{1st}) is equivalent to
\begin{equation} \label{1st-1}
    \frac{dE_{\rm S}}{dt}= P(t)+\sum_\alpha J_\alpha(t).
\end{equation}

This identity expresses the first law of thermodynamics for the driven 
quantum system: the variation of the internal energy equals the sum of the rate of work done 
by the driving forces and the heat currents exchanged with the baths. 
The slow-driving expansion preserves this balance order by order, ensuring the 
thermodynamic consistency of the approach.

\subsubsection{Entropy change}
Our aim is to relate the different components of the heat current and the power with the changes of the 
von Neumann entropy associated to the reduced density matrix of the system. Keeping terms of the density 
operator up to the first order in the slow-driving expansion, the entropy reads
\begin{equation}\label{entropy}
    S(t)= - k_B\; \mathrm{Tr}\;\bigg\{ \left(\rho^{(f)}(t)+\rho^{(1)}(t)\right) \ln \left(\rho^{(f)}(t)+\rho^{(1)}(t)\right)\bigg\},   
\end{equation}
which can be written as 
\begin{equation}\label{split}
     S(t)=S^{(f)}(t)+S^{(1)}(t),
\end{equation}
being
\begin{align}\label{sf1}
     S^{(f)} = - k_B\; \mathrm{Tr}\;\left\{\rho^{(f)} \ln{\rho^{(f)}} \right\}\;\;\;\;\;\;  \textmd{and}\;\;\; \;\;\;   S^{(1)} = - k_B\; \mathrm{Tr}\;\left\{ \rho^{(1)} \ln{\rho^{(f)}}\right\}.
\end{align}
Details are presented in Appendix~\ref{apent}, where we show that these components satisfy the following relations:
\begin{equation}\label{entrof}
     \frac{dS^{(f)}}{dt} = 
     \frac{1}{T}\sum_\alpha J^{(1)}_\alpha(t).
     %\bm \Lambda^1_\alpha(\bm X) \cdot  \dot{\bm X}
\end{equation}
and 
\begin{equation}\label{entro1}
    \frac{dS^{(1)}}{dt} =  \frac{1}{T} \left( \sum_\alpha J^{(2)}_\alpha + P^{(2)} \right) = \frac{1}{T} \frac{d E_S^{(1)}}{dt}.
\end{equation}
Equation~(\ref{entrof}) relates the heat fluxes produced by slow time-dependent changes in the system parameters to the variation of entropy along the sequence of quasi-static equilibrium states defined by the instantaneous parameter values. This corresponds to the linear-order contribution in ${\tau}^{-1}$ of the heat current, which is fully reversible and does not generate entropy over a complete driving cycle. 

On the other hand, Eq.~(\ref{entro1}) makes explicit the thermodynamic meaning of the second-order terms in the  slow-driving expansion.  
$P^{(2)}$ corresponds to work that does not remain stored in the system but is instead released as heat into the reservoirs through the second-order heat currents $J^{(2)}_\alpha$.  
The balance expressed in Eq.~(\ref{entro1}) ensures that this dissipated energy corresponds to the irreversible entropy production, which physically takes place in the reservoirs. Hence, $P^{(2)}$ and $J^{(2)}_\alpha$ constitute 
the two complementary manifestations of dissipation in the slow-driving regime.

\subsection{Cycles and geometric properties}

We now analyze the energy exchange over a complete driving cycle and its connection with the geometric aspects of the 
slow dynamics. To this end, we define the frozen system energy and its first-order correction as
\begin{equation}\label{defi}
    E_{\rm S}^{(f)}(t)={\rm Tr}\left[\rho^{(f)}(t) {\cal H}_{\rm S}(t) \right], \quad \quad E_{\rm S}^{(1)}(t)={\rm Tr}\left[\rho^{(1)}(t) {\cal H}_{\rm S}(t) \right].
\end{equation}
By substituting these definitions into Eq.~(\ref{1st-1}), and evaluating the corresponding terms 
order by order in the slow-driving expansion,  we can identify the contributions to the internal energy and heat 
flow associated with each order in ${\tau}^{-1}$.  
In particular, we focus on the net energy exchange over a full cycle of duration ${\tau}$, 
for which the control parameters return to their initial configuration, $\bm X({\tau})=\bm X(0)$.

\subsubsection{First order in the slow-driving expansion}
At the lowest order, the first-order correction to the system energy satisfies
\begin{equation}\label{heat-pow}
    \frac{dE_{\rm S}^{(f)}}{dt}= P^{(1)}(t)+\sum_\alpha J^{(1)}_\alpha(t),
\end{equation}
where $P^{(1)}(t)$ and $J^{(1)}_\alpha(t)$ denote, respectively, the first-order power delivered by the external driving and the heat current flowing from the reservoir $\alpha$ to the system.
Integrating~\eqref{heat-pow} over one period $\tau$, knowing that the system returns to the initial state,  
$E_{\rm S}^{(f)}({\tau})=E_{\rm S}^{(f)}(0)$, and using Eqs.~~\eqref{definitionsp} 
and \eqref{definitions}, we obtain
\begin{equation}
    0  = \int_0^{\tau} \frac{dE_{\rm S}^{(f)}}{dt} dt =\oint_C \text{Tr}\left[\partial_{\bm X}{\cal H}_{\rm S}(\bm X) \rho^{(f)}(\bm X)\right] \cdot  d\bm X+\sum_\alpha\oint_C \bm \Lambda^{(1)}_\alpha(\bm X) \cdot  d\bm X,
\end{equation}
where $C$ is the path traced by the driving protocol in parameter space $\bm X$.
The first integral is the work done by the conservative forces associated with the parametric dependence of the Hamiltonian and it vanishes over the cycle in parameter space. 
The other term defines the heat pumped between reservoirs because of the time-dependent driving at the central system,
\begin{equation} \label{pump}
Q^{(\rm pump)}_\alpha= \int_0^\tau J_\alpha^{(1)}(t)dt = \oint_{C} \bm \Lambda^{(1)}_\alpha(\bm X) \cdot  d\bm X. 
\end{equation}
Energy conservation at first order then implies
\begin{equation}\label{pumpzero}
  \sum_\alpha Q^{(\rm pump)}_\alpha = 0,
\end{equation}
which expresses the fact that the pumped heat corresponds to a redistribution of energy between reservoirs, without any net accumulation in the system after a full cycle.

We stress that $\bm \Lambda^{(1)}_\alpha(\bm X)$ depends only on the control parameters $\bm X$, but not on their velocities. Hence, $Q^{(\rm pump)}_\alpha$ is a purely geometric quantity akin to a Berry phase accumulated along the close contour $C$ in parameter space. Importantly, in order to have a non-vanishing value of this quantity, it is a necessary condition to have two driving parameters, which is a characteristic property of pumping in the adiabatic response regime \cite{Brouwer1998Oct,Moskalets2004May,Bhandari2020Oct}. 

% Using the Stokes theorem we can rewrite,
% \begin{equation}\label{rotor}
%     Q^{\rm pump}_\alpha= \iint_\Sigma \nabla\times\bm \Lambda^{(1)}_\alpha(\bm X) \; d^2X .
% \end{equation}

% In the case of a two-parameter protocol this rotor has one component, so it gives a simple picture of heat pumping as a function of $\bm X = [X_1,X_2] $. In the results we are going to show some of this rotors for different systems and couplings.

\subsubsection{Second order in the slow-driving expansion} \label{sec:2nd-order}
We now focus on the fulfillment of the first law, Eq. (\ref{1st-1}), at the next order in the slow-driving expansion. This leads to
\begin{equation}
    \frac{dE_{\rm S}^{(1)}}{dt}= P^{(2)}(t)+\sum_\alpha J^{(2)}_\alpha(t).
\end{equation}
Assuming that the system returns to its initial state after one cycle, $E_{\rm S}^{(1)}({\tau})=E_{\rm S}^{(1)}(0)$, we integrate the left-hand side over a period $\tau$ and 
substitute Eqs.~(\ref{definitionsp}) and (\ref{definitions}) into the right-hand side. This yields
\begin{eqnarray}\label{balance_order_2}
   0 = \int_0^{\tau}  \frac{dE_{\rm S}^{(1)}}{dt} dt & = 
  \sum_\alpha  \int_0^{\tau} \bigg(\dot{\bm X} \cdot \underline{\bm \Omega}^{(\lm{2})}_{\alpha}
\cdot \dot{\bm X} + \bm \Lambda_{\alpha}^{(2)} \cdot \ddot{\bm X} \bigg)\; dt +\int_0^{\tau}  
\dot{\bm X} \cdot \underline{\bm \Lambda} \cdot \dot{\bm X}\;dt.
\end{eqnarray}
The first terms define the second-order contribution to the total heat entering the system from each reservoir $\alpha$,
\begin{equation}\label{Q_2_alpha}
Q^{(2)}_\alpha =  \int_0^{\tau} \bigg(\dot{\bm X} \cdot \underline{\bm \Omega}^{(\lm{2})}_{\alpha}
\cdot \dot{\bm X} + \bm \Lambda^{(2)}_{\alpha} \cdot \ddot{\bm X} \bigg)\;dt.
\end{equation}
The last term of Eq.\eqref{balance_order_2} corresponds to the work generated by the time-dependent sources on the system, 
\begin{equation}\label{driving_work}
W^{(2)} =  \int_0^{\tau} \; 
\dot{\bm X} \cdot \underline{\bm \Lambda} \cdot \dot{\bm X}\;dt.
\end{equation}

This work can be lower bounded by a geometric quantity, the so-called thermodynamic length. Using the Cauchy-Schwarz inequality in Eq.~\eqref{driving_work}, we obtain
\begin{equation}\label{w-bound}
    W^{(2)} \ge \frac{1}{\tau}\left( \int_0^{\tau} \sqrt{\dot{\bm X} \cdot \underline{\bm \Lambda} 
    \cdot \dot{\bm X}}  \;dt\right)^2\; = \frac{1}{\tau}\left( \oint_C \sqrt{ d\bm X \cdot \underline{\bm \Lambda} \cdot d\bm X}  \right)^2 = \frac{\mathcal{L}^2}{{\tau}},
\end{equation}
where ${\cal L}$ is the thermodynamic length, which measures the length of the closed 
trajectory in the parameter space equipped with the metric tensor $\underline{\bm \Lambda}$.

When calculating $Q^{(2)}_\alpha$ in a cycle, it is possible to rewrite the term involving the "acceleration" $\ddot{\bm X}$,
\begin{equation}\label{hip_acc}
    \oint_C \Lambda_{\alpha,j}^{(2)} \: \ddot{X}_j \:dt = \left.\bm \Lambda^{(2)}_\alpha \cdot \dot{\bm X}\right|_{P_0}^{P_0} - \oint_C \frac{\partial \Lambda^{(2)}_{\alpha,j}}{\partial X_i} \dot{X}_i \dot{X}_j \:dt = \oint_C -\frac{\partial \Lambda^{(2)}_{\alpha,j}}{\partial X_i} \dot{X}_i \dot{X}_j \:dt,
\end{equation}
%\lm{Acá el Referee 1 nos pide que aclaremos bien todo lo asumido para tirar el termino de borde. Se asume continuidad de la velocidad en el punto inicial/final. Esto entra en un conflicto trivial con nuestro ciclo pizza, porque tranquilamente se podria redondear el vertice de la pizza (con un radio muy pequeño) y obtener una velocidad continua sin cambiar los resultados relevantes.}
\textcolor{red}{where $P_0$ denotes an (arbitrary) initial point of the cyclic protocol in the parameter space. $ \bm \Lambda^{(2)}_\alpha$ depends smoothly on the parameters and we focus on protocols
for which $\dot{\bm X}$ is well defined. Hence, the first term vanishes.  }
 This allows us to absorb the acceleration contribution into a velocity-quadratic form, which is more natural for interpreting dissipation.
Then, we define
\begin{equation}
     \lm{\tilde{\Omega}}^{(2)}_{\alpha,ij} = -\frac{\partial \Lambda^{(2)}_{\alpha,j}}{\partial X_i},
\end{equation}
so that the total second-order dissipative kernel in Eq.~\eqref{Q_2_alpha} is
\begin{equation}\label{omega_completo}
\underline{\bm \Omega}_\alpha = \underline{\bm \Omega}_\alpha^{(\lm{2})} + \lm{\tilde{\underline{\boldsymbol{\Omega}}}}_\alpha^{(2)}.
\end{equation}
Notice that this total kernel explicitly emerges when computing $Q^{(2)}_\alpha$ as the time integral 
of $J^{(2)}_\alpha$ over a full period $\tau$. However, for the second-order heat flux evaluated 
over an arbitrary time interval, the boundary contributions 
$\bm \Lambda^{(2)}_\alpha \cdot \dot{\bm X}$ in Eq.~\eqref{hip_acc} must also be taken 
into account.

Eq. (\ref{balance_order_2}) naturally leads to the identification of the total dissipated  heat entering the reservoirs,
\begin{equation}\label{qdiss}
    Q^{\rm{(diss)}}= - \sum_\alpha Q^{(2)}_\alpha=W^{(2)}.
\end{equation}
Physically, this expresses that all the extra work done by the driving forces beyond the conservative 
response is irreversibly transferred to the reservoirs. Integrating Eq.~(\ref{entro1}) over the cycle yields the 
same result, consistently identifying $Q^{\rm{(diss)}}/T$ with the total entropy generated by the slow driving.

\subsection{Landauer bound}

Here we consider Eqs. (\ref{entrof}) and (\ref{pump}) to find a bound for the maximum amount of heat that can be pumped per cycle
between the reservoirs.
We first notice that over a cycle, the total change of the frozen-state entropy vanishes
\begin{equation}
      0 = \int_0^{\tau} \frac{dS^{(f)}}{dt}\;dt =\oint_{C} \partial_{\bm X} S^{(f)}(\bm X) \cdot d \bm X,
\end{equation}
reflecting the fact that the system returns to its initial equilibrium state at the end of the cycle. 
We split the trajectory in the parameter space as $C=C^> + C^< $, 
where $\Delta{S}^{(f)}|_{C^>} >0$ corresponds to segments along which the entropy increases, 
and $\Delta{S}^{(f)}|_{C^<}=-\Delta{S}^{(f)}|_{C^>}$ to those along which it decreases.

Since $\rho^{(f)}$ is a thermal state, the largest possible entropy change along 
these paths is limited by the Landauer bound \cite{Landauer1961Jul,Landauer1988Oct}, 
\begin{equation}\label{sbound}
\left.\Delta{S}^{(f)}\right|_{C^>}^{\rm max}=k_B \ln N, \quad \quad \left.\Delta{S}^{(f)}\right|^{\rm min}_{C^<}=-k_B \ln N,
\end{equation}
which represents the difference between the entropy of an equiprobable state and that of a pure state in an $N$-level system.

For a system coupled to a single reservoir, Eq. (\ref{entrof}) is consistent with the fact  that, over a cycle, the net heat pumped between the system and the reservoir is zero. 
However, by spliting the trajectory on the parameter space as above, we can 
define $Q_{{\rm single},C^>}>0$ and $Q_{{\rm single},C^<}<0$
as the heat absorbed or released along each segment.
Using Eq. (\ref{entrof})  and Eq. (\ref{sbound}) we conclude
\begin{equation}\label{qbound}
Q^{\rm max}_{{\rm single},C^>}=k_B T \ln N, \quad \quad Q^{\rm min}_{{\rm single},C^<}=-k_B T\ln N,
\end{equation}
which clearly shows that the heat exchanged per segment is fundamentally bounded by the maximum entropy 
change of the system.

For a system with $M$ reservoirs, this generalize to
\begin{eqnarray}\label{qbound-1}
& & \sum_\alpha Q_{\alpha ,C^>}=\lm{T} \Delta{S}^{(f)}|_{C^>}\leq k_B \lm{T} \ln N,\quad \quad
\sum_\alpha Q_{\alpha,C^<}=-\lm{T}\Delta{S}^{(f)}|_{C^>}\geq -k_B \lm{T} \ln N,\nonumber \\
& & \sum_\alpha \left[Q_{\alpha ,C^>}+Q_{\alpha,C^<}\right]=0,
\end{eqnarray}
highlighting that while the heat pumped into individual reservoirs may fluctuate during the cycle, 
the total net exchange always vanishes, and each contribution remains constrained by 
the Landauer limit.

\subsection{Heat pumping and \textcolor{red}{interactions} }

In the general case of driven systems connected to several reservoirs, the analysis of the
Landauer bound for the entropy presented previously does not allow us to conclude on  the bound for the total pumped heat 
over a full cycle in a particular reservoir. 

For $N_{\rm q}$ non-interacting qubits, we can show that the net heat pumped over a cycle into any one of the coupled reservoirs is bounded as follows
\begin{equation}\label{bound-nonint}
\left|  Q_{\alpha}^{\rm (pump)}\right| \leq N_{\rm q} k_B T \ln 2 = k_B T \ln N \quad \textmd{for non-interacting qubits}.
\end{equation}
This inequality follows from the analytical solution of Lindblad equation
and the slow-driving expansion.
Details are presented in Appendix \ref{app:single-qubit} and \ref{app:many-qubits}. Substituting Eqs. (\ref{curr-adia-qubit}) and (\ref{ds-qubit}), we get
that the heat current exiting  the reservoir $\alpha$ at the first-order in the slow-driving expansion reads
\begin{equation} \label{j-s-nonint}
    J^{(1)}_\alpha(t)=\frac{  \Gamma_\alpha(t) }{\sum_\eta \Gamma_\eta(t)}   T \frac{dS^{(f)}}{dt} \quad \textmd{for non-interacting qubits},
\end{equation}
where $\Gamma_\alpha$(t), defined in the Appendix~\ref{app:single-qubit}, is a positive quantity.
From this equation, it is clear that the maximum heat that can be pumped into the reservoir $\alpha$ is $k_B T N_{\rm q} \ln 2$. \textcolor{red}{A concrete protocol that achieves the saturation of this bound} corresponds to  $\alpha$ connected during $C^>$, where the entropy increases, while all the other reservoirs are disconnected, and $\alpha$ is disconnected during $C^<$ where the entropy decreases while at least one of the remaining reservoirs is connected. 
Such a protocol corresponds to heat delivered from the reservoir $\alpha$ to the qubits along $C^>$, followed by the 
same amount of heat released from the qubits to the other reservoirs along $C^<$. 
The minimum bound $-k_B T N_{\rm q} \ln 2$ follows from an analogous reasoning. 

In Appendix~\ref{app:single-qubit} we first show Eq. (\ref{j-s-nonint}) for a single qubit. The bound Eq.~\eqref{bound-nonint} for $N_{\rm q}=1$ 
was previously found from a numerical analysis in Ref. \cite{TerrenAlonso2022Feb}. 
To show that the same bound is also valid for 
$N_{\rm q}$ non-interacting qubits, we notice that the state of this system can be expanded as a 
product state
\begin{equation}\label{N_qubits}
    \rho(t)=\frac{1}{\lm{2^{N_q}}} \left( \sigma^0 + \bm r_{1} \cdot \bm \sigma_1 \right)\otimes \cdots\otimes \left( \sigma^0 + \bm r_{N_q} \cdot \bm \sigma_{N_q} \right).
\end{equation}
Then, we can follow the same arguments as for a single qubit. Details are presented in  Appendix \ref{app:many-qubits}, where we find the stationary solution for the Lindblad equation by solving the linear set of equations for  
$\overline{\bm r}^{(f)}=\left( \overline{\bm r}^{(f)}_1, \ldots,\overline{\bm r}^{(f)}_{N_{\rm q}} \right)$, which is equivalent to $N_{\rm q}$ decoupled equations like Eq.~\eqref{ma-bloch}.

Importantly, for the case of interacting qubits, where the density operator cannot be expressed as a direct product, Eqs. (\ref{bound-nonint})
 and (\ref{j-s-nonint})
 cannot be proved to be valid. Furthermore, in numerical calculations for two interacting qubits, we show that 
heat can be pumped exceeding this bound. Hence, interactions and quantum correlations can be exploited to 
enhance heat pumping beyond the {\it Landauer-like} limit of independent qubits, revealing a clear link between 
\lm{correlation} and thermodynamic performance.

\section{Performance of slowly-driven many-qubit thermal machine operating between two thermal baths}

So far, we have considered thermal baths at the same temperature. 
Hence, we have  analyzed the mechanisms of pumping and dissipation, but not yet 
the mechanism of heat-work conversion. 
We now focus on the slow-driving cycle in a system in contact to two reservoirs at temperatures 
$T\pm \Delta T/2$. We assume a small temperature difference $\Delta T/T$, which justifies a linear-response treatment in this quantity. 

A configuration with a finite temperature bias and at least two slowly driving parameters operates as a thermal machine. 
At linear order, the heat flux and the generated power can be expressed as responses to the temperature bias and the rate of change of the driving parameters \cite{Ludovico2016Feb,Bhandari2020Oct}. 
With the convention that heat entering the system is positive, the total heat transferred to the cold reservoir and the net work done against the external sources over a cycle read
\begin{eqnarray}
Q^{\rm (net)} &=& Q_{\rm cold}^{\rm (pump)} + \kappa \; \frac{\Delta T}{T}, \nonumber \\
W^{\rm (net)} &=&-W^{(2)}+ Q_{\rm cold}^{\rm (pump)} \; \frac{\Delta T}{T}.
\end{eqnarray}
In the case of the heat, $Q^{(\rm pump)}_{\rm cold}$ denotes the geometrically pumped heat (see Eq.~\ref{pump}) --present even at $\Delta T = 0$-- while the additional term describes the net heat transferred in a cycle as a response to the temperature bias. $\kappa$ is proportional to the cycle-averaged thermal conductance between the reservoirs.
The first term in $W^{(\rm net)}$ is the work performed by the driving sources (see Eq.~\ref{driving_work}) and, as discussed in \ref{sec:2nd-order}, contributes 
to the dissipation. The key ingredient for the thermal machine operation is the second term, which describes the mechanism of heat-work conversion. 
Remarkably, this component  is proportional to the heat pumping and it is 
related to this mechanism by Onsager reciprocal relations \cite{Ludovico2016Feb,Bhandari2020Oct}. The sign of this term is defined by the protocol and it defines the operation of the machine as a heat engine for $Q_{\rm cold}^{\rm (pump)}>0$, or a refrigerator for $Q_{\rm cold}^{\rm (pump)}<0$. 

The heat-engine operation of these adiabatic thermal machines was analyzed in detail 
in  Ref.\cite{TerrenAlonso2022Feb}, and we summarize below the characterization of 
the performance presented in that paper. 
The net  generated power per cycle is $P=W^{\rm (net)}/\tau$. 
This quantity can be expressed in terms of geometric contributions as follows,
\begin{equation}\label{power-geo}
P=-\frac{L^2}{\tau^2}+\frac{A}{\tau}\frac{\Delta T}{T},
\end{equation}
where $A\equiv Q_{\rm cold}^{\rm (pump)}$ represents the geometrically pumped heat per cycle, 
which can be identified,  through the Stokes' theorem, as 
the oriented area enclosed by the cyclic trajectory in the space of control parameters, weighted by the Berry curvature  
$\nabla_{\bm X} \times \bm \Lambda^{(1)}_{\rm cold}(\bm X)$, with $\Lambda^{(1)}_{\rm cold}(\bm X)$ calculated from Eq. (\ref{lambda-al}). 
The coefficient $L^2 \equiv \tau \int_0^\tau \dot{\bm X}\cdot\underline{\bm \Lambda}(\bm X)\cdot \dot{\bm X}\,dt$   quantifies the dissipated work and can be interpreted as the squared length of the trajectory in a metric space defined by the kernel $\underline{\bm \Lambda}(\bm X)$.
The competition between these contributions determines an optimal driving period. 
From the maximization of $P$ with respect to the duration of the cycle,
it is found that the optimal duration is $\tau_D=2 T L^2/(A \Delta T)$, while the 
maximum achievable power is  
$P_{D}= 1/4 \left(A/L\right)^2 \left(\Delta T/T\right)^2$,
where $A$ and $L$ characterize, respectively, the magnitude of the pumped 
heat and of the dissipated power. 
This expression shows that the achievable power scales quadratically with the temperature bias 
and is determined solely by the geometric and metric properties of the driving protocol, 
although it still depends on the specific velocity profile through the evaluation of $L^2$.
However, taking into account Eq. (\ref{w-bound}), an optimal protocol regarding the velocity 
can be chosen, in order to guarantee a constant dissipation rate at every point of the trajectory, in which case, 
\begin{equation}\label{pot}
    P^{\rm max}= \frac{1}{4} \left(\frac{A}{\cal L}\right)^2 \left(\frac{\Delta T}{T}\right)^2.
\end{equation}
Therefore, the maximum power for the optimal duration of the cycle and for the optimal 
velocity of  the parameter change along the protocol is fully determined by the 
ratio between the area enclosed by the closed contour of the parameter space and the 
length of this contour.  As mentioned in Ref.\cite{TerrenAlonso2022Feb}, finding an optimal 
protocol is an open problem in geometry: the ``Cheeger problem'' \cite{Leonardi2015Nov}. 
However, it is possible to use Eq. (\ref{pot}) as a guide to find optimal parameters for selected
protocols.

\section{Results for two qubits}

We now present numerical results for a system composed of two qubits, each  
coupled to two thermal baths, labeled $\alpha=L,R$, both at the same temperature $T$.  
We consider the model with a Heisenberg exchange interaction between qubits,
as in Eq. (\ref{ham_sys}) for $N_{\rm q}=2$. 
The external magnetic fields acting on the qubits are time-dependent 
and given by $\bm B_1(t)=\left(B_x(t),0,B_z(t)\right)$ and $\bm B_2(t)=\eta\, \bm B_1(t)$, 
where the parameter $\eta$ controls the degree of asymmetry between the local drivings.
The Hamiltonian reads
\begin{equation}\label{ham_simplif}
\mathcal{H}_S(t) = - B_x(t) \; (S_1^x + \eta \; S_2^x ) -  B_z(t) \; (S_1^z + \eta \; S_2^z) + 
J \; \bm S_1 \cdot \bm S_2.
\end{equation}
In this configuration, the magnetic fields drive both qubits along the same trajectory 
in parameter space but with different amplitudes.  
The two thermal baths are coupled to the two-qubit system through the interaction Hamiltonian 
of Eq.~(\ref{c-bath}), with coupling Hermitian operators defined as 
\begin{equation} \label{contacts}
 \pi_L = S_1^x + b\, S_2^x, \qquad
 \pi_R = S_1^z + b\, S_2^z .
\end{equation}
The parameter $b$ introduces an asymmetry in the coupling strength between each qubit and the reservoirs. 
A global prefactor in the coupling would simply rescale the overall interaction rate 
and does not affect the results, unless it breaks the weak-coupling assumption.
Introducing the asymmetry factor $b$ in the coupling of the first qubit instead of the second one leads to equivalent qualitative behavior.

The numerical solution was obtained by vectorizing the density matrix and 
representing the Lindbladian and the dissipator operators as matrices acting on this 
vectorized space. As a validation of our present Lindbladian framework, we have verified 
that our results 
reproduce those reported in Ref.~\cite{TerrenAlonso2022Feb}.  
The latter were obtained from a master equation derived perturbatively 
within the Keldysh non-equilibrium Green's function formalism~\cite{Bhandari2021Jul}.

\begin{figure}
    \centering
    \begin{subfigure}{0.48\textwidth}
        \centering
        \includegraphics[width=\linewidth]{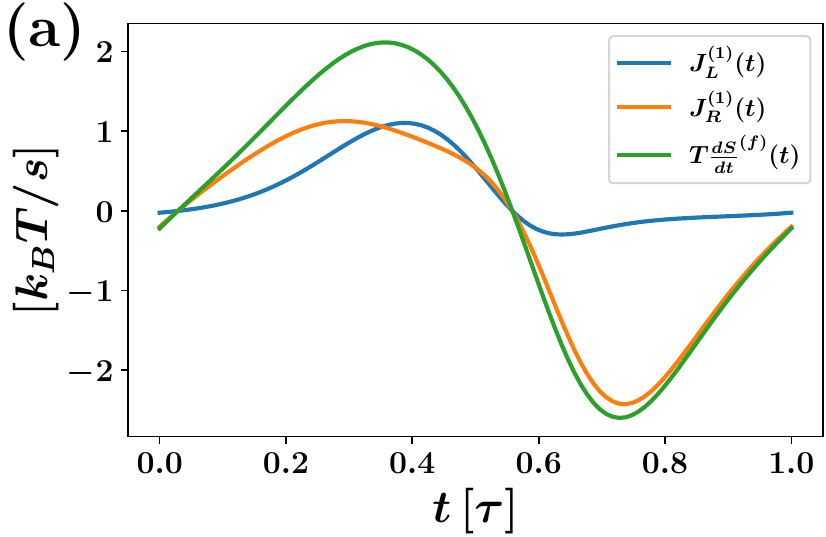}\label{consistency_check1}
        %\label{consistency_check1}
    \end{subfigure}\hfill
    \begin{subfigure}{0.48\textwidth}
        \centering
        \includegraphics[width=\linewidth]{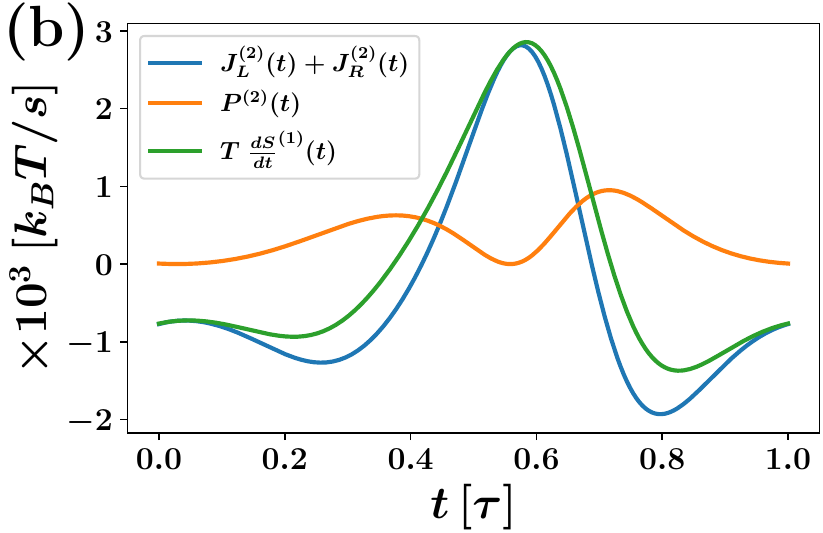}
        %\label{consistency_check2}
    \end{subfigure}
     \caption{Benchmark of the entropy--energy balance for the two-qubit system coupled to two thermal reservoirs 
 $L$ and $R$ at equal temperature. The external fields follow the driving protocol of Eq.~(\ref{proto}) with $B_0 = k_B T$, 
 as illustrated in Fig.~\ref{fig:J0}.  Panel (a): the time-dependent first-order heat currents entering the 
 system from reservoirs $L$ and $R$ (blue and orange lines) sum to the instantaneous rate of change of the system entropy 
 (green line) (see Eq.~\ref{entrof}).  Panel (b): the sum of the second-order heat currents (blue) and the 
 driving power (orange)  reproduces the second-order entropy variation (green) (see Eq.~\ref{entro1}).
        For the numerical calculations we take $J=0$ and $\eta=1.2$ in~\eqref{ham_simplif}, 
        contact asymmetry $b=2$ in Eq.~(\ref{contacts}), and reservoir parameters $g_L=g_R=0.002$, 
        $\omega_C=120\,k_B T$ in Eq.~(\ref{gammas}).}
            \label{fig:consistencias}
\end{figure}

We start by presenting results that explicitly demonstrate the thermodynamic consistency 
of the present approach. These results are shown in 
Fig.~\ref{fig:consistencias}, where the heat currents flowing from the reservoirs into the system 
are shown for the driving protocol 
\begin{equation}\label{proto}
    B_x(t)=B_0\left[1+0.5 \cos{(2\pi t)}\right], \quad\quad\quad B_z(t)=B_0\left[0.5+0.25 \sin{(2\pi t)}\right].
\end{equation}
which defines a periodic cycle of duration $\tau=1$ in the chosen units. 
Figure~\ref{fig:consistencias}(a) corresponds to Eq.~\eqref{entrof}.
Because the couplings and the driving protocol are asymmetric, the heat currents 
from the left and right reservoirs differ in magnitude.
Nevertheless, their sum reproduces at all times the rate of change of the system 
entropy, $dS^{(f)}(t)/dt$, in full agreement with Eq.~(\ref{entrof}). 
Furthermore, the integral of this quantity over a full period vanishes, 
as expected for a cyclic operation, while the time integrals of the individual heat currents satisfy 
$\int_0^\tau\! dt\, J^{(1)}_L(t) = -\int_0^\tau\! dt\, J^{(1)}_R(t) = Q^{(\mathrm{pump})}_L$.

We can also verify that  $d S^{(1)}(t)/dt$ integrates zero over the cycle, 
while the integral of the other plotted quantities  
are $Q^{(2)}_L = 15.5 \;k_B T$, $Q^{(2)}_R = -364.9\; k_B T $ and $W^{(2)} = 349.4\; k_B T$, which verifies 
Eq.~\eqref{balance_order_2} and Eq. (\ref{qdiss}). 

% \lmc{¿No habría que comentar acerca de que está entrando calos al sistema desde el reservorio izquierdo, cuando
% se esperaria solo calor hacia los reservorios?}

Results corresponding to the benchmark for Eq.~(\ref{entro1}) are presented in Fig.~\ref{fig:consistencias}(b). 
The figure shows the total second-order heat flux entering the system from both reservoirs, the power developed by 
the driving sources, and the corresponding second-order entropy variation. 
We find that Eq.~(\ref{entro1}) is satisfied at every time, confirming the internal consistency of the 
slow-driving expansion. 

Overall, the numerical benchmarks show that the present formalism correctly captures both the 
reversible and irreversible components of the dynamics.
\bigskip

\subsection{Heat pump and Landauer bound}

We now discuss  the heat pumping $Q^{(\rm pump)}_\alpha$. Using the Stokes' theorem in the expression~\eqref{pump}  
of the pumped heat, we can express it as the flux of a rotor through the surface $\Sigma$, defined as the surface enclosed 
by the  external parameter protocol contour $C$,
\begin{equation}\label{rotor}
    Q^{(\rm pump)}_L= -\;Q^{(\rm pump)}_R=\iint_\Sigma \left[ \nabla_{\bm X}\times\bm \Lambda^{(1)}_\alpha(\bm X)\right]\cdot \; d \bm \Sigma.
\end{equation}
In the two-parameter protocol considered here, the rotor reduces to a single component, 
providing a simple picture of heat pumping as a function of $\bm X = (B_x, B_z)$.

\begin{figure}[!htb]
    \centering
    \begin{subfigure}{0.48\textwidth}
        \centering
        \includegraphics[width=\linewidth]{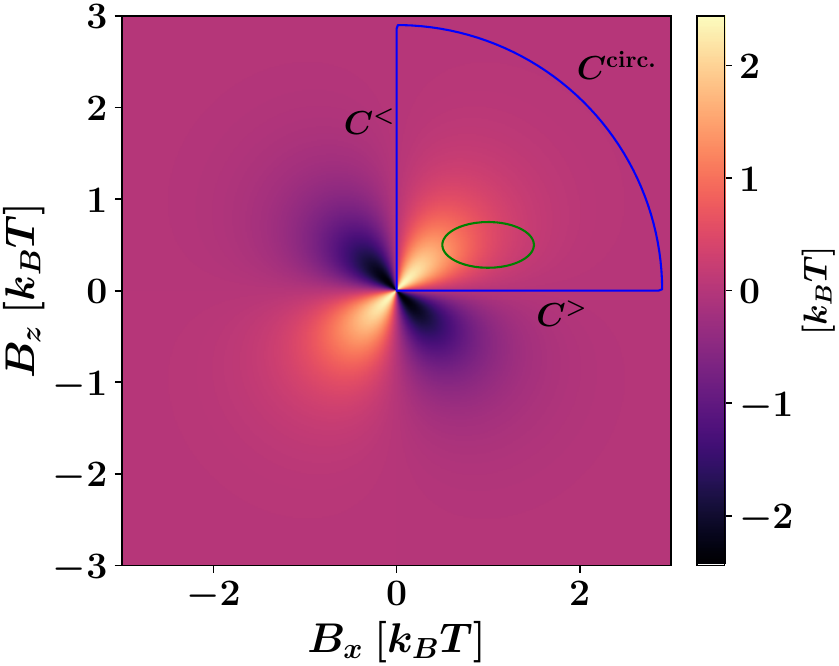}
        \caption{$\nabla\times\bm \Lambda^{(1)}_L$ with $J=0$.}
        \label{fig:J0}
    \end{subfigure}\hfill
    \begin{subfigure}{0.48\textwidth}
        \centering
        \includegraphics[width=\linewidth]{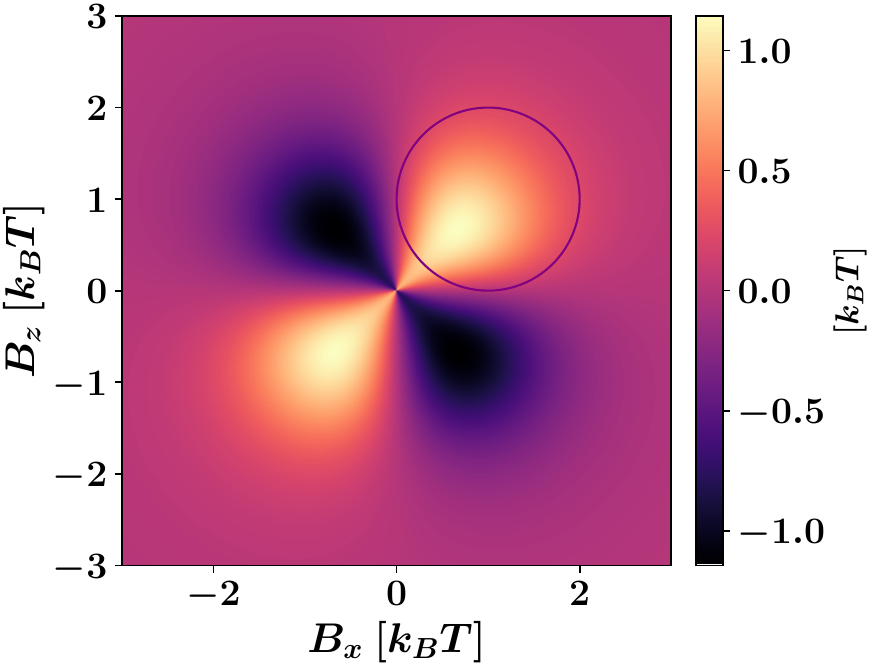}
        \caption{$\nabla\times\bm \Lambda^{(1)}_L$ with $J=2 \; k_B T$.}
        \label{fig:J2asym}
    \end{subfigure}
%    \begin{subfigure}{0.33\textwidth}
%        \centering
%        \includegraphics[width=\linewidth]{pictures_2qubits/rotorL_J=2.png}
%        \caption{}
%        \label{fig:J2}
%    \end{subfigure}\hfill
    \caption{
Rotor field in the parameter space $\bm X=(B_x,B_z)$ for different configurations of the two-qubit system. The closed paths $C$ indicate the driving cycles for which the corresponding results are presented in the following figures.  Panel~ (a) shows the non-interacting case, while panel~(b) illustrates the case 
with inter-qubit coupling $J = 2\,k_B T$. All other parameters are the same as in Fig.~\ref{fig:consistencias}.}
\label{fig:rotors}
\end{figure}

In Fig.~\ref{fig:rotors} we display the rotor for several configurations of the two-qubit system and indicate the driving cycles $C$ for which the corresponding results are discussed below.
Fig. \ref{fig:J0} shows the results for the non-interacting two-qubit system ($J=0$). We can identify the same features as those reported in Ref. \cite{TerrenAlonso2022Feb} for a single qubit with comparable couplings to the reservoirs. 
In particular, the driving protocol that encloses a single quadrant in parameter space leads to the largest pumped heat. 
Hence, to reach a bound for the pumped heat, we choose a trajectory analogous to the 
circular sector shown in Figure \ref{fig:J0}, with a large radius encopassing 
the whole quadrant. 
This circular sector can be divided into three parts: one horizontal straight line denoted $C^{>}$, a circular arc, and a vertical straight line $C^{<}$ as shown in Figure \ref{fig:J0}. 
The heat pumped over a cycle is calculated integrating as shown in Eq.~\eqref{pump}, while the change in entropy is calculated from Eq. (\ref{entrof}).
In $C^{>}$ we obtain $Q_{R,C^>} = \lm{T} \Delta S^{(f)}|_{C^{>}} = 1.3862\; k_B T \simeq 2 k_B T \ln 2$ 
and $Q_L=0$. For the circular arc we find that the heat interchange, as well as the change of entropy is negligible and vanishes exactly as the radius of the circular sector tends to infinity. For the vertical straight line $C^{<}$ we find 
$Q_{L,C^<} = \lm{T} \Delta S^{(f)}|_{C^<} = - 1.3862\; k_B T \simeq 2 k_B T \ln 2$ and $Q_R=0$.

These results can be understood from the pattern of energy exchange between the 
system and the reservoirs along each segment of the cycle. 
Along the path $C^{>}$ ($B_z=0$), the state of both qubits are polarized in the $x$ direction of the Bloch sphere. Hence, they can exchange energy with the right reservoir, which is coupled through the operator $\pi_R$ introduced in Eq.~\eqref{contacts}. 
This operator involves $S^z_j$ terms, which enable transitions between the different eigenstates of $S^x_j$. During this process the energy levels separate, and therefore, the system absorbs energy from the right reservoir. In the circular path, the  exchange of energy between the system and the reservoirs is negligible when the radius is large enough. This is because the energy levels do not change during this stroke. In $C^{<}$ ($B_x=0$), the system evolves only in contact with the left reservoir and 
is decoupled from the right one. The reason is that, along this path, 
the system is polarized along $z$ in the Bloch sphere. 
Hence, they can exchange energy 
with the reservoir contacted through the operator $\pi_L$ of Eq. (\ref{contacts}) which involves 
$S^x_j$ operators. Along this path, the energy difference between the levels decreases, implying an energy flux from the system to the left reservoir. 
In summary, an amount of heat $Q_{R,C^>}$ enters the system from the right reservoir into the system 
along $C^{>}$ and it is released to the left reservoir along  $C^{<}$. 
We can identify the total pumped heat of the cycle as $Q^{(\rm pump)} = Q_{R,C^>}=-Q_{L,C^>} \simeq 2\;k_B T \ln 2$.

We now discuss the results shown in Fig. \ref{fig:J2asym}, corresponding to two interacting qubits described by Eq.~\eqref{ham_simplif} with $J=2$, and coupling operators Eq.~\eqref{contacts} with asymmetry factor $b=2$. 

We notice similar features as in the non-interacting case. In particular, it is clear that the circular sector enclosing one of the quadrants is also the optimal protocol in this case. Hence, we analyze the heat pumped for the same circular sector indicated in the left panel of Fig.~\ref{fig:J2asym}.  
For the present case, the numerical evaluation 
casts $Q_{L,C^{>}} = -0.6848\; k_B T$, $Q_{R,C^{>}}=1.6028\; k_B T$, and  $\Delta S^{(f)}|_{C^{>}} = 0.9180\; k_B$. We note that the system couples with both reservoirs due to the exchange interaction $J$ between the two qubits. 
For the circular path, the heat exchanged and the entropy change are negligible. 
For the rest of the protocol, we get $Q_{L,C^{<}} = -1.6028 \;k_B T$, $Q_{R,C^{<}}=0.6848 \;k_B T$, $\Delta S^{(f)}|_{C^{<}} = -0.9180 \;k_B$. 
It is important to notice that these results satisfy
\begin{eqnarray}
Q_{L,C^{>}} + Q_{R,C^{>}} &=& \;\;\; T\,\Delta S^{(f)}\big|_{C^{>}} \leq \;\;\; k_B T \ln 4, \nonumber \\
Q_{L,C^{<}} + Q_{R,C^{<}} &=& -T\,\Delta S^{(f)}\big|_{C^{<}} \geq - k_B T \ln 4,
\end{eqnarray}
in full consistency with Eq.~\eqref{qbound-1}. Nevertheless, 
the total heat pumped between reservoirs over the full cycle, given by Eq.~\eqref{pump}, is $Q^{\rm (pump)}_{L/R} = \pm 2.2876 \;k_B T$, which is larger 
than the Landauer bound expressed by Eq. (\ref{bound-nonint}),
satisfied by non-interacting qubits.

Remarkably, these results show that such a bound for the heat pump can be exceeded as a consequence of the \lm{correlation} between qubits introduced by the interaction $J$. In Appendix \ref{ent}, we analyze how the interaction between qubits originates an extra component in the heat current, 
in comparison with the result for the product state describing the non-interacting case. 
Such a contribution also depends on the type of coupling between the qubits and the reservoirs 
and does not necessarily enhance the heat pump. In fact, by modifying the couplings in the 
previous case to consider the symmetric case ($b=1$ in Eq.~\ref{contacts}), 
we get the following results for the same protocol and the same value of $J$: 
  $Q_{L,C^>} = 0$, $Q_{R,C^>} =0.9180\; k_B T$, $\Delta S^{(f)}|_{C^>} = 0.9180 \; k_B$, and $Q_{L,C^<} = -0.9180 \;k_B T$, $Q_{R,C^<} =0$, $\Delta S^{(f)}|_{C^<} = -0.9180 \;k_B$,
  while the contribution over the arc is negligible, as in the previous cases. 
  In this situation the heat pump is lower than $k_B T \ln 4$, in contrast to the results previously discussed.

%\begin{equation}
%\renewcommand{\arraystretch}{1.5} % Ajustar la altura de las filas
%\setlength{\arraycolsep}{10pt} % Ajustar el espacio entre las columnas
%\begin{array}{c|c|c}
%\mathrm{Horizontal\:straight\:line} &  \mathrm{Circular\:arc} & \mathrm{Vertical\:straight\:line} \\
%\hline
%Q^{\rm pump}_L = -0.6848 & Q^{\rm pump}_L = 0 & Q^{\rm pump}_L = -1.6028\\
%Q^{\rm pump}_R = 1.6028 & Q^{\rm pump}_R = 0 & Q^{\rm pump}_R = 0.6848\\
%\Delta S^{(f)} = 0.9180  & \Delta S = 0 & \Delta S = - 0.9180\\
%\end{array}
%\label{table_pumps3}
%\end{equation}

%\begin{equation}
%\renewcommand{\arraystretch}{1.5} % Ajustar la altura de las filas
%\setlength{\arraycolsep}{10pt} % Ajustar el espacio entre las columnas
%\begin{array}{c|c|c}
%\mathrm{Horizontal\:straight\:line} &  \mathrm{Circular\:arc} & \mathrm{Vertical\:straight\:line} \\
%\hline
%Q_L = 0 & Q_L = 0 & Q_L = -0.9180\\
%Q_R = 0.9180 & Q_R = 0 & Q_R = 0\\
%\Delta S^{(f)} = 0.9180  & \Delta S = 0 & \Delta S = -0.9180\\
%\hline
%& Q^{\rm pump}_L = -0.19189 & Q^{\rm pump}_L = 0 & Q^{\rm pump}_L = -0.94093\\
%a_x=1,a_z=0,b_x=0,b_z=1&Q^{\rm pump}_R = 1.10991 & Q^{\rm pump}_R = 0 & Q^{\rm pump}_R = 0.02291\\
%&\Delta S = 0.91801  & \Delta S = 0 & \Delta S = -0.91801\\
%\end{array}
%\label{table_pumps2}
%\end{equation}

% s1x-s2z  A_L=-1.13282

%EllipticCycle([1,0.5],0.5,0.25,0)

\subsection{Second order energy balance and dissipation}

We now turn to analyze the second-order component of the heat current. As already explained in Sec \ref{sec:2nd-order}, this is related to the  dissipated power in the form of heat. We notice that in the calculation of the net quantities over a cycle presented in Eq.~\eqref{balance_order_2}, 
 the matrices $\Omega_{L,ij}$, $\Omega_{R,ij}$ in \eqref{omega_completo} and $\Lambda_{ij}$ are contracted with the symmetric tensor $\dot{X}_i \dot{X}_j$. Therefore, only the symmetric part of these matrices  contributes to the balance. 
 In the two-qubit system  described by Eq.~\eqref{ham_simplif} we have found that
\begin{equation}\label{balance2_matrices}
\underline{\bm \Omega}^{(s)}_{L}(\bm X) + \underline{\bm \Omega}^{(s)}_{R}(\bm X) = -\underline{\bm \Lambda}^{(s)}(\bm X),
\end{equation}
where the superscript $(s)$ indicates that these are the symmetric parts of the matrices.
This result is valid for any set of parameters $\{ J,\eta \}$ of the Hamiltonian \eqref{ham_simplif} and factor $b$ in Eq.~\eqref{contacts}, and, when it is integrated over a full cycle, constitutes a verification of the second order energy balance expressed by Eqs.~\eqref{balance_order_2} and \eqref{qdiss}, which are valid for any cycle.

\begin{figure}[!htb]
    \centering
    \begin{subfigure}{0.315\textwidth}
        \centering
        \includegraphics[width=\linewidth]{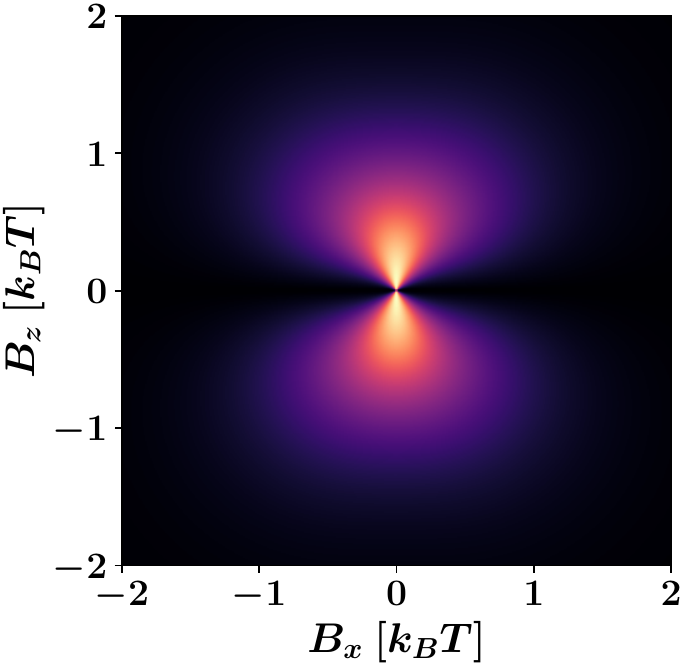}
        \caption{$\mathrm{MaxEigenvalue}(-\underline{\Omega}_L)$}
        \label{OmL}
    \end{subfigure}\hfill
    \begin{subfigure}{0.29\textwidth}
        \centering
        \includegraphics[width=\linewidth]{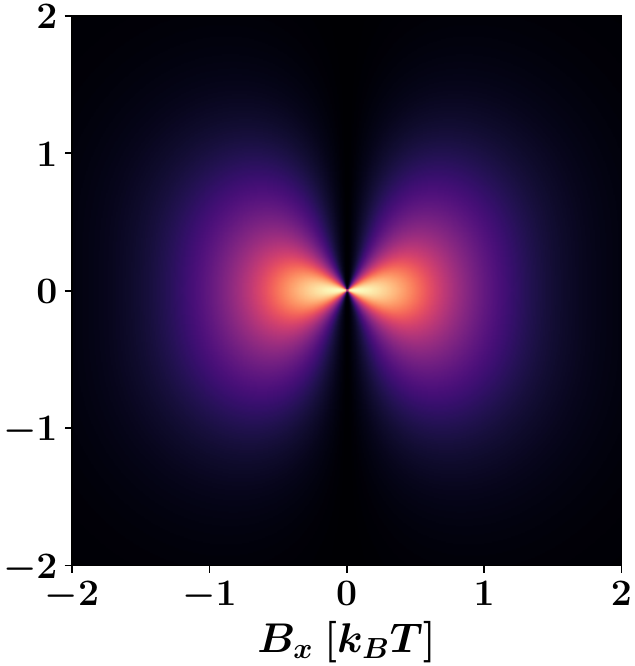}
        \caption{$\mathrm{MaxEigenvalue}(-\underline{\Omega}_R)$}
        \label{OmR}
    \end{subfigure}\hfill
    \begin{subfigure}{0.36\textwidth}
        \centering
        \includegraphics[width=\linewidth]{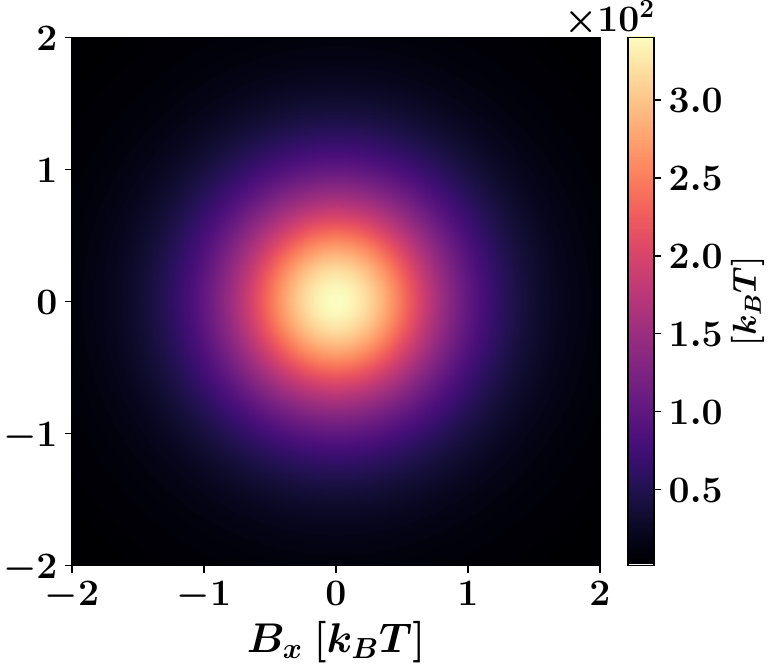}
        \caption{$\mathrm{MaxEigenvalue}(\underline{\Lambda})$}
        \label{Lambda_J=0}
    \end{subfigure}
    \caption{
    Maximum eigenvalue of the symmetric parts of  $-\underline{\Omega}_L$, $-\underline{\Omega}_R$, 
    and  $\underline{\Lambda}$ in the parameter space $(B_x,B_z)$. Panels (a) and (b) show the asymmetric distribution of dissipation between reservoirs due to the coupling asymmetry, while $\underline{\Lambda}$ displays a symmetric pattern.
    System parameters as in Fig.~\ref{fig:consistencias} with $J=0$.}
    \label{lambda_and_omegas_J=0}
\end{figure}

The largest eigenvalue of the symmetric part of the matrices $\underline{\bm{\Omega}}_\alpha$ and 
$\underline{\bm \Lambda}$ indicates the maximum local dissipation rate achievable for a given driving speed. 
It thus provides a simple way to quantify both the intensity and the anisotropy of dissipation across the parameter 
space. These results are shown in Fig.~\ref{lambda_and_omegas_J=0}. The behavior observed in panels (a) and (b) reflects the asymmetric distribution of the dissipated heat between the two reservoirs, resulting from the asymmetry in the couplings. In contrast, the maximum eigenvalue of $\underline{\bm{\Lambda}}$ displays a symmetric pattern in parameter space. It is important to note that the eigenvalues of $\underline{\bm{\Omega}}_L + \underline{\bm{\Omega}}_R$ are not equal to the sum of the eigenvalues of each matrix, since the matrices $\underline{\bm{\Omega}}_\alpha$ do not commute.
Nevertheless, we can infer that the result of computing the integrals in 
Eqs. \eqref{Q_2_alpha} and \eqref{driving_work} with these matrices, in general,
implies an asymmetric distribution of the dissipated energy $W^{(2)}$ between the two reservoirs.

\begin{figure}[!htb]
    \centering
    \begin{subfigure}{0.42\textwidth}
        \centering
        \includegraphics[width=\linewidth]{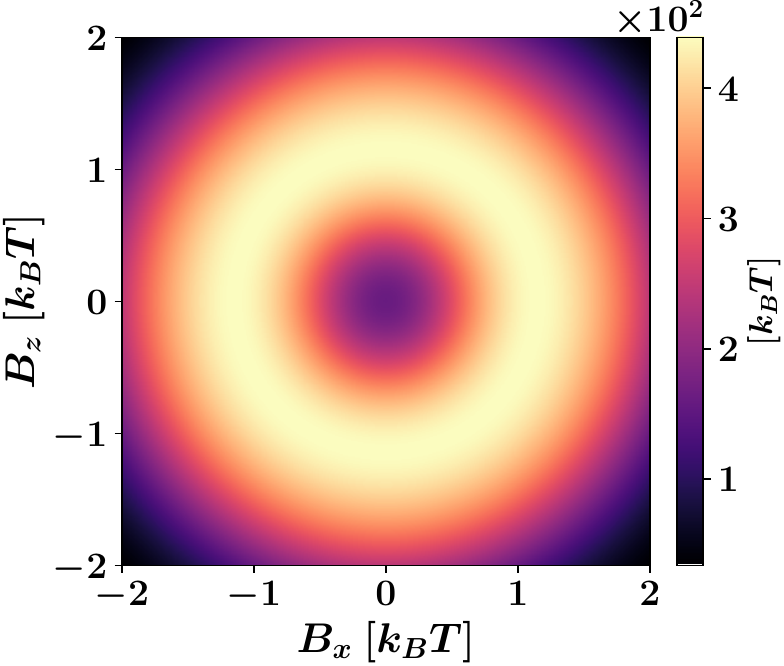}
        \caption{$J=1 \; k_B T$}
        \label{Lambda_mat_J=1}
    \end{subfigure}\hfill
    \begin{subfigure}{0.42\textwidth}
        \centering
        \includegraphics[width=\linewidth]{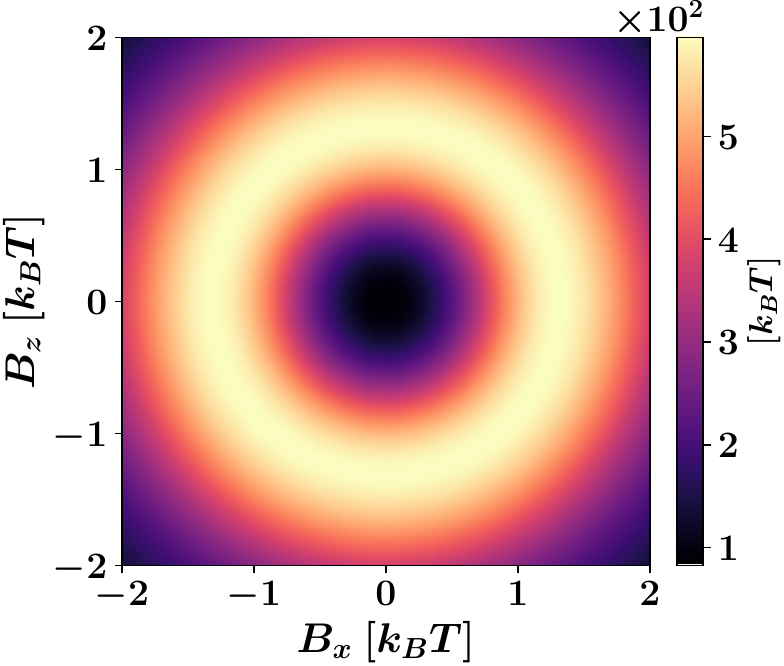}
        \caption{$J=2 \; k_B T$}
        \label{Lambda_mat_J=2}
    \end{subfigure}
\caption{Maximum eigenvalue of the symmetric part of matrix $\underline{\Lambda}$ for interacting qubits 
with $J = 1\; k_B T$ (a) and $J = 2\; k_B T$ (b). The interaction modifies both the magnitude and distribution 
of dissipation compared to the non-interacting case. Other parameters as in Fig.~\ref{fig:consistencias}.}
    \label{Lambdas_con_interaccion}
\end{figure}

The effect of the interaction $J$ between qubits in the dissipated energy is illustrated in Fig.~\ref{Lambdas_con_interaccion}.  Compared with the non-interacting case shown in Fig.~\ref{Lambda_J=0}, 
the interacting system develops a ring of enhanced intensity in the $(B_x,B_z)$ plane, whose 
radius and overall magnitude increase with $J$, while the dissipation decreases in the central region around $B_x = B_z = 0$. 
As a consequence, the interaction redistributes the dissipation across parameter space: depending on the 
path of the protocol, it may either suppress or enhance the net dissipated energy.

\subsection{Figure of merit of the heat engine}

%On the other hand, looking at the rotors \ref{fig:J0} and \ref{fig:J2asym} we guess that there must be protocols where the figure of merit \eqref{figure_merit} is better in the interacting qubits system, because it is characterized by the quotient $(\frac{A}{L})^2$.

%\begin{figure}
%    \centering
%    \begin{subfigure}{0.25\textwidth}
%        \centering
%        \includegraphics[width=\linewidth]{discusiones/Figure_merit_r=0.1_J=0.png}
%        \caption{$J=0$, $a=1,b=2$}
        %\label{OmL}
%    \end{subfigure}\hfill
%    \begin{subfigure}{0.25\textwidth}
%        \centering
%        \includegraphics[width=\linewidth]{discusiones/Figure_merit_r=0.1_J=2.png}
%        \caption{$J=2$, $a=1,b=2$}
        %\label{OmR}
%    \end{subfigure}\hfill
%    \begin{subfigure}{0.25\textwidth}
%        \centering
%        \includegraphics[width=\linewidth]{discusiones/Figure_merit_r=0.1_J=3.415.png}
%        \caption{$J=3.415,a=1,b=1.35$}
        %\label{Lambda_J=2}
%    \end{subfigure}
%    \begin{subfigure}{0.25\textwidth}
%        \centering
%        \includegraphics[width=\linewidth]{discusiones/Figure_merit_r=0.1_J=5_f2=1.35.png}
%        \caption{$J=5$, $a=1,b=1.35$}
        %\label{Lambda_J=2}
%    \end{subfigure}
%    \caption{\centering
%        Figuras de merito en circulos de radio $r=0.1$ centrados en cada punto. Da que la interaccion la empeora en todos los casos}
    %\label{lambda_and_omegas_J=2}
%\end{figure}

We now discuss the behavior of the figure of merit $A^2/\mathcal{L}^2$ that quantifies the optimal power of the two 
driven qubits operating as a heat engine when they are coupled to the $L,R$ reservoirs with a temperature 
bias $\Delta T$, assuming $\Delta T/T\ll 1$. 

\begin{figure}[!htb]
    \centering
    \includegraphics[scale=0.6]{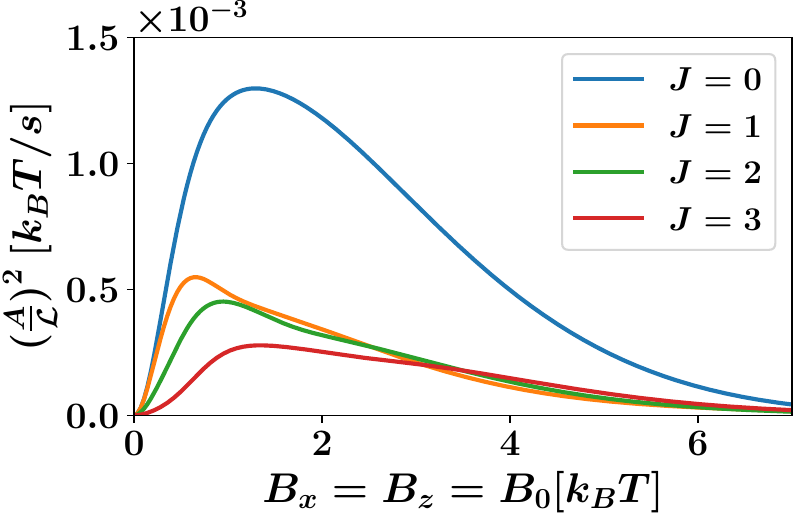}
    \caption{Figures of merit in circles defined by the protocol Eq. (\ref{proto1}). The system and coupling parameters are the same as for Figure \ref{fig:consistencias}, with different values of $J$. }
    \label{figures_of_merit}
\end{figure}

Results are shown in Fig.~\ref{figures_of_merit} for protocols defined as circular trajectories in parameter space that describe a circle of radius $B_0$ centered at $(B_0,B_0)$ in the $(B_x,B_z)$ plane, tangent to both coordinate axes, and given by
\begin{equation}\label{proto1}
B_x(t)=B_0\!\left(1+\cos 2\pi t\right), \qquad
B_z(t)=B_0\!\left(1+\sin 2\pi t\right).
\end{equation}

Notice that $A$ is the geometrical area enclosed by the cycle path $C$ in the parameter space, weighted by the 
rotors shown in Fig.~\ref{fig:rotors}, while the lengths that set the dissipation bounds, $\mathcal{L}^2$, are 
determined by the kernels displayed in Figs.~\ref{Lambda_J=0}, \ref{Lambda_mat_J=1}, and \ref{Lambda_mat_J=2}. 
All these quantities are symmetric with respect to the identity line $B_x = B_z$, so the circular protocols provide 
an overview of the behavior of the figure of merit within the whole quadrant. For this class of protocols, we find that the interaction $J$ does not enhance the power performance of the heat engine, in spite of the fact that we found an enhancement of the pumped heat.

%\begin{figure}[!htb]
%    \centering
%    \includegraphics[scale=0.48]{pictures_2qubits/A_L_HD.png}
%    \caption{Figures of merit for circles of radius $r_c=0.1$ centered in the position $(r,r)$ in the identity line. The system and coupling parameters are the same as for Figure \ref{fig:consistencias} with different values of $J$.}
%    \label{figures_of_merit}
%\end{figure}

%% FIGURA CON LOS OMEGAS Y LAMBDA E J=2
%\begin{figure}
%    \centering
%    \begin{subfigure}{0.33\textwidth}
%        \centering
%        \includegraphics[width=\linewidth]{pictures_2qubits/-OmegaL_J=2_asym.png}
%        \caption{$\mathrm{MaxEigenvalue}(-\underline{\Omega}_L)$}
%        \label{OmL}
%    \end{subfigure}\hfill
%    \begin{subfigure}{0.33\textwidth}
%        \centering
%        \includegraphics[width=\linewidth]{pictures_2qubits/-OmegaR_J=2_asym.png}
%        \caption{$\mathrm{MaxEigenvalue}(-\underline{\Omega}_R)$}
%        \label{OmR}
%    \end{subfigure}\hfill
%    \begin{subfigure}{0.33\textwidth}
%        \centering
%        \includegraphics[width=\linewidth]{pictures_2qubits/Lambda_J=2_asym.png}
%        \caption{$\mathrm{MaxEigenvalue}(\underline{\Lambda})$}
%        \label{Lambda_J=2}
%    \end{subfigure}
%    \caption{\centering
%        $J=2$ in $\mathcal{H}_S$ and $a_x=a_z=1,b_x=b_z=2$ in the coupling operators.}
%    \label{lambda_and_omegas_J=2}
%\end{figure}

\subsection{Role of the number of qubits and the system–bath couplings}

To finalize, we compare the behavior of the different quantities analyzed previously for the two-qubit system, with the results obtained for a single driven qubit. We also analyze the effects introduced by the details of the couplings to the reservoirs. 

\begin{figure}[!htb]
\centering
\begin{minipage}[t]{0.48\textwidth}
    \includegraphics[width=\textwidth]{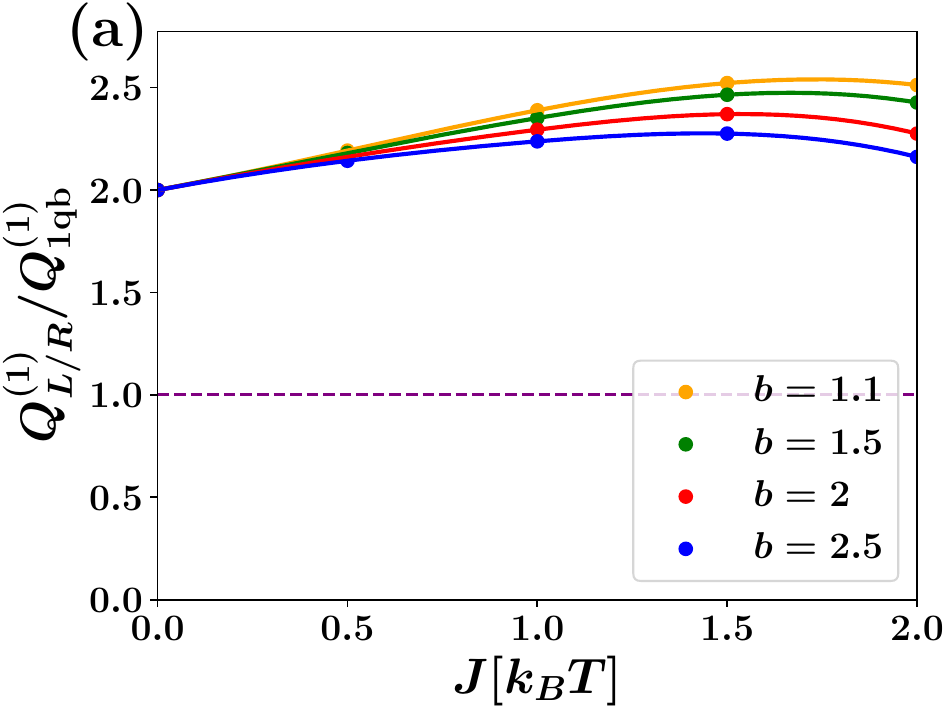}
\end{minipage}%
%\hspace*{-0.38em}%
\begin{minipage}[t]{0.48\textwidth}
    \includegraphics[width=\textwidth]{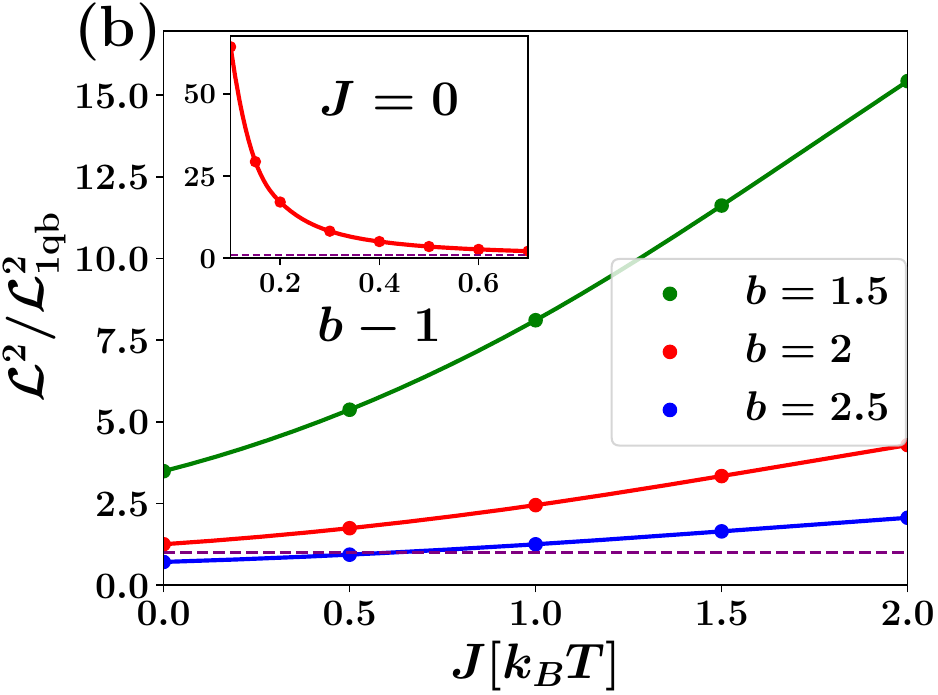}
\end{minipage}%
\caption{%
        Heat pumping in (a) and minimum dissipation in (b) as a functions of the interaction parameter $J$ for different parameter $b$, calculated for the protocol Eq. (\ref{proto1}) with $r=1,\; B_0=k_B T$
        and plotted in \ref{fig:J2asym}. 
        The respective values for a single qubit system $Q_{\rm 1qb}^{\rm (pump)}= 0.527 k_B T,\mathcal{L}^2_{\mathrm{1qb}}/\tau= 712 [k_B T]$ are shown as a reference in dashed purple lines.
        Inset: minimum dissipation $\mathcal{L}^2$ is shown for the non-correlated qubit system $J=0$ divided by the corresponding value for a single qubit $\mathcal{L}^2_{\mathrm{1qb}}$ as a function of the parameter $b$. Other parameters 
         are the same as in Figure \ref{fig:consistencias}.
        }
\label{cosas_vs_J}
\end{figure}

In Fig.~\ref{cosas_vs_J}(a), we show values of the heat pumping $Q^{(1)}_{L/R}$ 
as a function of the interaction parameter $J$ for the protocol of Eq. (\ref{proto1}) plotted in Fig.~\ref{fig:J2asym}. We notice that the pumped heat  in the case of two non-interacting qubits ($J=0$) duplicates the value obtained for a single qubit (presented in dotted lines in  the figure).
Results are shown for different values of the parameter $b$, representing the asymmetry in the couplings of each qubit to the reservoirs,  as indicated in Eq.~\eqref{contacts}. 
We see that for this protocol, the interaction between qubits enhances the pumping and this effect is stronger for the smaller asymmetries shown in the figure.

In Fig.~\ref{cosas_vs_J}(b) we show the minimum dissipation $\mathcal{L}^2$ as a function of the interaction parameter $J$ for the same protocol and asymmetry parameter $b$ as in  panel (a). 
For comparison, we also plot the results for a single qubit, $\mathcal{L}^2_{\mathrm{1qb}}$, in dashed lines. 
In all the cases, increasing $J$ leads to a larger value of the minimum dissipation, 
with a strong dependence  on the asymmetry parameter $b$. 
Remarkably, for strong asymmetries (see plot with $b=2.5$), 
the minimum dissipation per qubit can be lower than in the case of a single qubit. 
In the Inset of Fig.~\ref{cosas_vs_J} we analyze this dependence in more detail in the case of  non-interacting qubits. 
We see a significant increment of the minimum dissipation system as $b\to 1$ (corresponding to symmetric coupling), while it decreases significantly as the asymmetry increases.

\section{Summary and conclusions}

We have analyzed in detail the thermodynamic consistency of the Lindbladian slow-driving approach 
introduced in Ref. \cite{Cavina2017Aug}, which provides  a general framework for describing 
thermal machines composed of many interacting subsystems weakly coupled to reservoirs. 
This formalism, valid in the linear-response regime with small temperature biases~\cite{Bhandari2020Oct}, 
has been extensively employed to investigate dissipation and finite-time thermodynamic cycles~\cite{Scandi2019Oct,Abiuso2020Mar,Rolandi2023Nov}. 

In the present paper, we extend the analysis to the behavior of heat pumping between reservoirs at 
equal temperatures, which also defines the heat-work conversion mechanism in the thermal machine operating 
with a thermal bias. For the single qubit case, a similar analysis has been followed \cite{TerrenAlonso2022Feb}, 
on the basis of a master equation derived from perturbation theory and non-equilibrium Green's 
functions \cite{Bhandari2021Jul}. 
In this work, we recover those results and also extend the analysis on the basis of the Lindbladian 
approach, to calculate the heat current up to second order in the slow-driving expansion, 
which is a key step to analyze the thermodynamic consistency.  
In addition,  we have shown analytically that the lowest order contribution 
to the pumped heat (the geometric component) is bounded by the Landauer limit, $k_B T \ln N_{\rm q}$, 
for a system of $N_q$ non-interacting qubits. This bound can be surpassed when 
qubit–qubit interactions are introduced, revealing the role of correlations in enhancing the 
heat-pumping capability of driven quantum systems.

We have also presented numerical benchmarks for a system of two coupled qubits, explicitly 
confirming the energy and entropy balance derived from the slow-driving expansion. 
Our results allowed us to characterize how the qubit-qubit interaction and the asymmetry in the reservoir 
couplings influence the pumped heat, the dissipation, and the power generated by the heat engine. 

Our result for the two-qubit system is consistent with a scaling of the optimal generated power of the heat engine $\propto N_q^2$, 
irrespectively of the interaction between qubits but upon optimization of the protocol and 
implementing asymmetric couplings between each qubit and the reservoirs. 
The latter scaling was reported in Ref. \cite{tajima2021superconducting, Kamimura2022May, tajima2} for a system of interacting qubits 
operating as a four-stroke heat engine and "superradiant" couplings to the reservoirs. 
The analysis of the two-qubit system is non-conclusive with respect to the advantage introduced 
by the interaction between qubits in the generated power of the heat engine. The numerical results for circular protocols suggest that 
it is detrimental. However, we notice 
that the role of the interactions in both; the dissipation and the figure
of merit of the heat engine, strongly depends on the choice of the protocol. 

The approach presented here can be easily extended to analyze other interacting quantum platforms, 
like quantum dots and nanomechanical systems under driving. \textcolor{red}{Many results have been already 
reported on quantum systems in strong coupling with the reservoirs under 
 slow-driving regime, among others \cite{Ludovico2014Apr,Ludovico2016Feb,Ludovico2016Jul,Ludovico2018Jan,Strasberg2019Oct,Strasberg2021Aug,Dou2020May,Deghi2024Sep}.
 Furthermore, a quantization in the heat current in units of $k_B T \log(2)/\tau$ has been found in quantum dots weakly coupled to the reservoirs under periodic driving. Most of these results have been compiled in
 \cite{Acciai2025Oct}.
 Nevertheless, many aspects on the role of the many-body interactions in the problem of energy conversion efficiency and the interplay between coherence  and dissipation remain to be explored. 
 Another interest direction for future investigations is the analysis of the response in the temperature bias beyond the linear terms in combination with the slow-driving expansion.}

\section*{Acknowledgements}
We thank Rosario Fazio for many stimulating conversations, as well as Marti Perarnau-Llobet for reading the manuscript and providing useful comments.

\paragraph{Funding information}
We acknowledge support from CONICET and FONCyT through PICT 2020-A-03661, Argentina in the early stages of this work (LA and GC).

\begin{appendix}

\numberwithin{equation}{section}

\section{Derivation of the Lindblad equation}\label{Lindblad_derivation}

We summarize the key steps leading to the Lindblad equation for the frozen Hamiltonian, following 
Ref.~\cite{Breuer2002}. The complete Hamiltonian is $\mathcal{H} = \mathcal{H}_{\rm S} + \sum_{\alpha} \mathcal{H}_\alpha + \sum_{\alpha} V_{\alpha}$. $\mathcal{H}_{\rm S}$ describes the system, $\mathcal{H}_\alpha$ the $\alpha$ reservoir, and 
${\cal V}_{\alpha} = g_\alpha \pi_{\alpha}B_\alpha,$ the coupling between the system and each reservoir, 
with $\pi_{\alpha}$ and $B_\alpha$ Hermitian operators acting on the system and on the reservoir $\alpha,$ respectively.
Taking $V = \sum_{\alpha} {\cal V}_{\alpha}$ as the interaction, we integrate once 
the von Neumann equation for the full-system density operator 
in the interaction representation $\frac{d\tilde{\rho}(t)}{dt} = - i \left[\tilde{\cal V}(t),\tilde{\rho}(t)\right],$
and reinsert it into the equation to obtain the second-order expression
\begin{equation}\label{ec_mov_orden1}
 \frac{d\tilde{\rho}(t)}{dt} = - i \left[\tilde{\cal V}(t),\tilde{\rho}(0)\right] - 
 \int_0^t d\tau \left[\tilde{\cal V}(t),\left[\tilde{\cal V}(\tau),\tilde{\rho}(\tau)\right]\right]. 
\end{equation}
Assuming weak system–reservoir coupling and keeping terms up to $g_\alpha^2$ in Eq.~\eqref{ec_mov_orden1} (Born approximation), 
we replace $\tilde{\rho}(\tau) = \tilde{\rho}_s(\tau) \otimes \rho_b$, with the reservoirs remaining in a stationary state 
at all times. Tracing out the bath degrees of freedom, the first term in \eqref{ec_mov_orden1} 
reduces to $\langle B_\alpha\rangle$, which may be set to zero, so the reduced density matrix of the system obeys
\begin{equation}
 \frac{d\tilde{\rho}_s(t)}{dt}   =  \sum_{\alpha}g^2_\alpha \int_0^t ds  
 \tilde{\pi}_\alpha(t)\bigg[{\cal B}_{\alpha}(-s)\tilde{\rho}_s(t-s)\tilde{\pi}_\alpha(t-s) - 
 {\cal B}_{\alpha}(s) \tilde{\pi}_\alpha(t-s)\tilde{\rho}_s(t-s)
 \bigg] + {\rm H.c.}
 \end{equation}
Here, we define the bath correlators ${\cal B}_{\alpha}(s)\delta_{\alpha\beta} = \langle \tilde{B}_\alpha(s)\tilde{B}_\beta(0)\rangle$. Next, we apply the Markov approximation, assuming 
that ${\cal B}_{\alpha}$  decays rapidly on a time scale much shorter than that of the system-reservoir dynamics. 
In this situation, we approximate $\tilde{\rho}_s(t-s) \simeq \tilde{\rho}_s(t),$ making the master equation local in time, and extend the upper limit of the integral to infinity. 

To obtain the Lindblad form, we still need to invoke the secular approximation. 
We begin by performing the spectral decomposition of the operators $\pi(t)$ as 
$\pi_\alpha =  \sum_{\omega} \pi_{\alpha \omega}$ where $\omega$ are the Bohr frequencies of the system and 
\begin{equation}
    \pi_{\alpha\omega} = \sum_{\substack{lm \\ \varepsilon_m-\varepsilon_l = \omega}}
    \xi_{lm}^{\alpha}|l\rangle\langle m|,
    \end{equation}
with $\xi_{lm}^{\alpha} = \langle l|\pi_\alpha|m\rangle$.
Since $\pi_{\alpha\omega}$ are eigenoperators of ${\cal H}_{\rm S}$, obeying $\left[{\cal H}_{\rm S},\pi_{\alpha\omega}\right] = -\omega\pi_{\alpha\omega}$, their interaction representations are $\tilde{\pi}_{\alpha\omega} = e^{-i\omega t}\pi_{\alpha\omega}$. 
Thus, the coupling operators become $\tilde{\pi}_\alpha(t) = \sum_{\omega} e^{-i\omega t}\pi_{\alpha,\omega} = \sum_{\omega}e^{i\omega t}\pi^\dagger_{\alpha,\omega}.$ Inserting this into the master equation yields 
\begin{equation}
\frac{d\tilde{\rho}_s(t)}{dt}  = \sum_{\omega \omega'} e^{i(\omega' -\omega)t}\sum_{\alpha} g^2_\alpha 
\bigg[\Gamma^*_{\alpha}(\omega')\pi_{\alpha \omega} \tilde{\rho}_s(t) \pi_{\alpha\omega'}^\dagger 
- \Gamma_\alpha(\omega)\pi^\dagger_{\alpha \omega'} \pi_{\alpha \omega} \tilde{\rho}_s(t)\bigg] +
\rm{H.c.}
   \label{eq:master1}
\end{equation}
where the bath correlator is defined as $\Gamma_{\alpha}(\omega) \equiv \int_0^{\infty} {\cal B}_{\alpha}(s)e^{i\omega s}ds 
=  \int_0^{\infty} \langle \tilde{B}_\alpha(s)\tilde{B}_\alpha\rangle e^{i\omega s}ds.$

If the characteristic time $\tau_R$ of the system-reservoir dynamics is much larger than the time scale $1/|\omega-\omega'|$ of the isolated system dynamics, then the terms with $\omega \neq \omega'$ oscillate rapidly and average out, 
leaving only the $\omega = \omega'$ contributions. This is the secular (or rotating-wave) approximation.   
After applying it and adding the unitary evolution term, we obtain the Lindblad equation for $\tilde{\rho}_s$ 
(up to a Lamb-shift correction of the $\cal H_{\rm S}$ spectrum):
\begin{equation}
 \frac{d\tilde{\rho}_s(t)}{dt}   =   -i\left[{\cal H}_{\rm S},\tilde{\rho}_s(t)\right] +  \sum_{\alpha} {\cal D}_{\alpha} \left[\tilde{\rho}_s(t)\right],
   \label{eq:master6}
\end{equation}
where the dissipator associated with each reservoir is defined  as 
\begin{equation}
{\cal D}_\alpha(\rho) =  \sum_\omega \bigg(L_{\alpha, \omega}
\rho L_{\alpha \omega}^\dagger -\frac{1}{2}\left\{L^\dagger_{\alpha \omega} 
L_{\alpha \omega}, \rho\right\}\bigg), 
 \label{eq:disipadores}
\end{equation}
in terms of the jump operators $L_{\alpha\omega}  =  g_\alpha \sqrt{\gamma_\alpha(\omega)} \pi_{\alpha,\omega}$, 
with $\gamma_\alpha$ the transition rate function 
\begin{equation}
 \gamma_\alpha(\omega) = \Gamma_{\alpha}(\omega) + \Gamma^*_\alpha(\omega) = 
 \int_{-\infty}^{\infty} \langle \tilde{B}_\alpha(s)\tilde{B}_\alpha\rangle e^{i\omega s} ds.
\end{equation}

%%%%%%%%%%%%%%%%%%%%%%%%%%%%%%%%%%%%%%%%%%%%%%%%%%%%%%%%%%%%%%%%%%%%%%%%%%%%%%
%%%%%%%%%%%%%%%%%%%%%%%%%%%%%%%%%%%%%%%%%%%%%%%%%%%%%%%%%%%%%%%%%%%%%%%%%%%%%%

\section{Derivation of the heat current}\label{appendix_heat_current}

Here we show that, by following the same steps used in the derivation of the Lindblad equation 
in Appendix~\ref{Lindblad_derivation}, we obtain Eq.~\eqref{heatalf} for the heat current.  
The heat current entering the system from reservoir $\alpha$ is defined as
\begin{equation}
 J_{\alpha} = -\frac{d}{dt} \langle\mathcal{H}_\alpha\rangle = -\frac{d}{dt} \mathrm{Tr} \left( \tilde{\rho}(t)\mathcal{H}_{\alpha} \right) = -\mathrm{Tr} \left(\frac{d\tilde\rho(t)}{dt} \mathcal{H}_\alpha\right)
\end{equation}
since $dH_{\alpha}/dt = 0$. Using the von Neumann equation in the interaction picture and the cyclic property of the trace, we obtain
\begin{eqnarray}
 J_\alpha(t)  = i g_{\alpha}\mathrm{Tr} \left(\tilde{\rho}(t)\tilde{\pi}_{\alpha}(t)\left[\mathcal{H}_{\alpha},\tilde{B}_{\alpha}(t)\right]\right).
\end{eqnarray}

Next, we insert the relation  
$\tilde{\rho}(t) = \tilde{\rho}(0) - i \int_0^t d\tau \left[\tilde{\cal V}(\tau),\tilde{\rho}_s(\tau)\right],$  
obtained by integrating the von Neumann equation once:
\begin{eqnarray}\label{current_eq1}
 J_{\alpha}(t)\! = \!i g_\alpha \mathrm{Tr} \left(\tilde{\rho}(0)\tilde{\pi}_{\alpha}(t)\left[\mathcal{H}_{\alpha},\tilde{B}_{\alpha}(t)\right]\right)\! +\! g_\alpha \int_0^t d\tau \; \mathrm{Tr} \left(\left[\tilde{V}(\tau),\tilde{\rho}(\tau)\right]
 \tilde{\pi}_{\alpha}(t)\left[\mathcal{H}_{\alpha},
 \tilde{B}_{\alpha}(t)\right]\right).
\end{eqnarray}

Under the Born approximation, we replace $\tilde{\rho}(\tau)$ inside the integral by 
$\tilde{\rho}_s(\tau)\otimes\rho_b$, assuming that the reservoirs remain in a stationary state. 
The first term vanishes for bosonic reservoirs, since it involves $\langle \left[\mathcal{H}_{\alpha},\tilde{B}_{\alpha}(t)\right]\rangle \propto \langle \tilde{B}_\alpha(t)\rangle$.  
Considering that the reservoirs are independent, the second term in Eq.~\eqref{current_eq1} 
becomes
\begin{eqnarray}
J_{\alpha}(t) =  g^2_{\alpha}\int_0^t d\tau \; \mathrm{Tr} \left(\left[\tilde{\pi}_{\alpha}(\tau)\tilde{B}_{\alpha}(\tau), \tilde{\rho}_s(\tau) \otimes \rho_b\right]\tilde{\pi}_{\alpha}(t)\left[\mathcal{H}_{\alpha},\tilde{B}_{\alpha}(t)\right]\right).
\end{eqnarray}

We now apply the Markov approximation (Appendix~\ref{Lindblad_derivation}):
\begin{eqnarray}
 J_{\alpha}(t) = g^2_{\alpha} \int_0^{\infty} d\tau\, \mathrm{Tr} \!\left([\tilde{\pi}_{\alpha}(t-\tau)\tilde{B}_{\alpha}(t-\tau),
 \tilde{\rho}_s(t) \otimes \rho_b]\,\tilde{\pi}_{\alpha}(t)[\mathcal{H}_{\alpha},\tilde{B}_{\alpha}(t)]\right).
\end{eqnarray}

Using the spectral decomposition $\tilde{\pi}_{\alpha}(t) = \sum_{\omega}\tilde{\pi}_{\alpha\omega}(t)$ (Appendix~\ref{Lindblad_derivation}), and performing the secular approximation to remove rapidly oscillating terms, we obtain
\begin{eqnarray}
 J_\alpha(t) = -\sum_{\omega}\omega g_\alpha^2 \Gamma_\alpha(\omega)\mathrm{Tr}_s\left(\pi_{\alpha,\omega}\tilde{\rho}_s(t)\pi^\dagger_{\alpha,\omega}\right) -
  \sum_{\omega}\omega g_\alpha^2 \Gamma^*_\alpha(\omega) \mathrm{Tr}_s\left(\tilde{\rho}_s(t)\pi^\dagger_{\alpha,\omega}
 \pi_{\alpha,\omega}\right).
\end{eqnarray}

Finally, decomposing the bath correlator $\Gamma$ into its real and imaginary parts, we arrive at
\begin{eqnarray}
 J_\alpha(t) = -g_\alpha^2\sum_{\omega}\omega  \gamma_\alpha(\omega) \mathrm{Tr}_s\left(\tilde{\rho}_s(t)\pi^\dagger_{\alpha,\omega}
 \pi_{\alpha,\omega}\right).
 \label{eq:jota1}
\end{eqnarray}

We now demonstrate that this expression for the heat current can also be written as Eq.~\eqref{heatalf}.  
To that end, we start from Eq.~\eqref{heatalf} and show that it reproduces Eq.~\eqref{eq:jota1}. Using the dissipators~\eqref{eq:disipadores} obtained from the Lindblad equation, we write
\begin{eqnarray}
 J_\alpha(t) & = & \mathrm{Tr}_s\bigg({\cal D}_\alpha\left[\tilde{\rho}_s(t)\right] {\cal H}_{\rm S}\bigg) = \nonumber \\
   & = & g_\alpha^2 \sum_{\omega} \gamma_\alpha(\omega)  
 \left[\mathrm{Tr}_s\bigg(\pi_{\alpha,\omega}\tilde{\rho}_s(t)\pi^\dagger_{\alpha,\omega}{\cal H}_{\rm S}\bigg)- \mathrm{Tr}_s\bigg(\tilde{\rho}_s(t)\pi^\dagger_{\alpha,\omega} \pi_{\alpha,\omega}{\cal H}_{\rm S}\bigg)\right].
 \label{eq:jota2}
\end{eqnarray}

To obtain this expression, we used that $[{\cal H}_{\rm S},\pi^\dagger_{\alpha \omega}\pi_{\alpha \omega}] = 0$.  
In the first term, we substitute $\pi^\dagger {\cal H}_{\rm S}$ by ${\cal H}_{\rm S}\pi^\dagger - \omega\pi^\dagger$, and since ${\cal H}_{\rm S}$ commutes with $\pi^\dagger \pi$, Eq.~\eqref{eq:jota1} is recovered.  

Thus, we conclude that
\begin{equation}
 J_\alpha(t) = -\frac{d}{dt}\langle \mathcal{H}_\alpha\rangle(t) = \mathrm{Tr}_s\bigg({\cal D}_{\alpha}\left[\tilde{\rho}_s(t)\right]{\cal H}_{\rm S}\bigg).
\end{equation}

%%%%%%%%%%%%%%%%%%%%%%%%%%%%%%%%%%%%%%%%%%%%%%%%%%%%%%%%%%%%%%%%%%%%%%%%%%%%%%%%%%

\section{Relations between change of entropy and the heat current}\label{apent}

In this Appendix we compute the first terms of the slow-driving expansion of the 
system entropy in powers of $\tau^{-1}$, where $\tau$ is the period of the protocol 
in the control parameter~$\bm X$. We start from the von Neumann entropy 
of the system,
\begin{equation}
    S = -k_B \mathrm{Tr} \left[\rho \ln \rho\right],
\end{equation}
and substitute the expansion $\lm{\rho = \rho^{(f)} + \rho^{(1)} + \cdots }$, where
$\rho^{(f)}$ is the thermal state of the frozen Hamiltonian $\mathcal{H}_S$ and
$\rho^{(1)} \equiv \tau^{-1}\tilde\rho^{(1)}$ denotes the first-order (adiabatic) correction.
This gives
\begin{equation}
\lm{
S \simeq -k_B\,\mathrm{Tr}\!\left\{
\left(\rho^{(f)} + \rho^{(1)}\right)
\ln\!\left( \rho^{(f)} \left[ \mathbb{I} + \rho^{(f)-1}\rho^{(1)} \right] \right)
\right\} .
}
\end{equation}
\lm{Separating the logarithm of a product into a sum and using the Taylor series of $\ln{(1+x)}$ we get}
%\begin{equation}
%S \simeq -k_B\,\mathrm{Tr}\!\left\{
%\left(\rho^{(f)} + \rho^{(1)}\right) \left( \ln{\rho^{(f)}} + \ln\!\left( \mathbb{I} + \rho^{(f)-1} \rho^{(1)}\right)  \right) \right\}
%\end{equation}
\begin{equation}
\lm{
S \simeq -k_B\,\mathrm{Tr}\!\left\{
\left(\rho^{(f)} + \rho^{(1)}\right) \left( \ln{\rho^{(f)}} +  \sum_{n=1} \frac{(-1)^{n+1}}{n}  \left( \rho^{(f)-1} \rho^{(1)} \right)^n \right) \right\} ,
}
\end{equation}
distributing
\begin{align}
S \simeq -k_B\,\mathrm{Tr}\!\left\{  \rho^{(f)} \ln{\rho^{(f)}} + \rho^{(1)} \ln{\rho^{(f)}} 
+ \rho^{(f)} \sum_{n=1}^{\infty} \frac{(-1)^{n+1}}{n}  \left( \rho^{(f)-1} \rho^{(1)} \right)^n \nonumber \right.\\
\left. + \rho^{(1)} \sum_{n=1}^{\infty} \frac{(-1)^{n+1}}{n}  \left( \rho^{(f)-1} \rho^{(1)} \right)^n \!\right\} .
\end{align}
\lm{We are only interested in frozen and first order in $1/\tau$, so we can discard the last terms, which are quadratic and of higher orders.}
\begin{equation}
    S \simeq -k_B\,\mathrm{Tr}\!\left\{  \rho^{(f)} \ln{\rho^{(f)}} + \rho^{(1)} \ln{\rho^{(f)}} + \rho^{(f)} \left( \rho^{(f)-1} \rho^{(1)} + \rho^{(f)} \sum_{n=2} \frac{(-1)^{n+1}}{n}  \left( \rho^{(f)-1} \rho^{(1)} \right)^n \right) \right\} \nonumber
\end{equation}
Again, discarding the higher order terms and noting that $\rho^{(1)}$ is traceless we finally get
\begin{equation}
\lm{
S \simeq -k_B\,\mathrm{Tr}\!\left\{  \rho^{(f)} \ln{\rho^{(f)}} \right\} - k_B\mathrm{Tr}\!\left\{ \rho^{(1)} \ln{\rho^{(f)}}  \right\}
}.
\end{equation}
The first term corresponds to $S^{(f)}$ and the second one to $S^{(1)}$, as given in
Eq.~(\ref{sf1}). 
Using the explicit form of the frozen density matrix,
$\rho^{(f)} = e^{-\beta \mathcal{H}_{\rm S}}/Z^{(f)}$ with
$Z^{(f)}=\mathrm{Tr}\left[e^{-\beta \mathcal{H}_{\rm S}}\right]$, the time derivative of
$S^{(f)}$ can be expressed as
\begin{eqnarray} \label{sdot}
    \frac{dS^{(f)}}{dt}
    &=& -k_B \frac{d}{dt} \mathrm{Tr}\left[\rho^{(f)}\left(-\beta \mathcal{H}_{\rm S}
    - \ln Z^{(f)}\right)\right] = \nonumber\\
    & = & \frac{1}{T}\,\mathrm{Tr}\left[
    \frac{d\rho^{(f)}}{dt}\mathcal{H}_{\rm S}+ \rho^{(f)} \frac{d\mathcal{H}_{\rm S}}{dt}\right]
    + k_B\,\frac{\dot Z^{(f)}}{Z^{(f)}} = \frac{1}{T}\;\mathrm{Tr}\left[\frac{d\rho^{(f)}}{dt}\mathcal{H}_{\rm S}\right].
\end{eqnarray}
Using the slow-driving expansion of the Lindblad equation (see Eq.~\ref{slow-exp}), 
this leads directly to Eq.~(\ref{entrof}):
\begin{equation}\label{demo_balance1}
     \frac{dS^{(f)}}{dt} = \frac{1}{T} \; \mathrm{Tr} \bigg\{{\cal L}_f \left[\rho^{(1)}\right]\;{\cal H}_{\rm S}\bigg\} =     \frac{1}{T}  \sum_\alpha \mathrm{Tr}\bigg\{{\cal D}_\alpha[{\rho}^{(1)}]\;{\cal H}_{\rm S} \bigg\} =  \frac{1}{T}\;\sum_\alpha\; J^{(1)}_\alpha(t).
\end{equation}
Next, we rewrite $S^{(1)}$ as
\begin{align}
    S^{(1)} &= - k_B\; \mathrm{Tr}\left[\rho^{(1)} \ln{\rho^{(f)}}\right] = - k_B\; \mathrm{Tr}\left[ \rho^{(1)} \left(-\beta \mathcal{H}_S - \ln Z^{(f)} \right)  \right] = \notag\\
    & =  \frac{1}{T} \;\mathrm{Tr}\left[\rho^{(1)}\; \mathcal{H}_S \right] + 
    k_B \ln Z^{(f)} \;\mathrm{Tr}\; \rho^{(1)} = \frac{1}{T}\; \mathrm{Tr}\left[\rho^{(1)} \mathcal{H}_S\right],
\end{align}
and evaluate its time derivative,
\begin{equation}
    \frac{dS^{(1)}}{dt} = \frac{1}{T} \; \mathrm{Tr}\left[ \frac{d\rho^{(1)}}{dt}  \;
    \mathcal{H}_S + \rho^{(1)} \frac{d \mathcal{H}_S}{dt}\right].
\end{equation}
Using $d \rho^{(1)}/dt = \sum_\alpha \mathcal{D}_\alpha[\rho^{(2)}]$, we find 
Eq.~\eqref{entro1}:
\begin{equation}
\frac{dS^{(1)}}{dt} = \frac{1}{T}  \mathrm{Tr}\bigg\{ \sum_\alpha 
\mathcal{D}_\alpha[\rho^{(2)}] \mathcal{H}_S + \rho^{(1)} 
\frac{d \mathcal{H}_S}{dt}\bigg\} = 
\frac{1}{T} \left( \sum_\alpha J^{(2)}_\alpha + P^{(2)} \right) = \frac{1}{T} \frac{d E_S^{(1)}}{dt}.
\end{equation}

\section{Exact analytical solution of Lindblad equation for a single driven qubit}\label{app:single-qubit}

We study a single-qubit system driven by a magnetic field $\bm B(t) = B \bm n_B(t)$, with Hamiltonian 
\begin{equation}
{\cal H}_S(\bm B) = \bm B \cdot \bm \sigma,
\end{equation}
being $\bm \sigma = \left(\sigma^x, \sigma^y, \sigma^z\right)$ the Pauli matrices. The instantaneous eigenstates are the two states $|\pm\rangle$ of the Bloch sphere, parallel and antiparallel to the 
$\bm n_B$ direction, with energies $\epsilon_{\pm} = \pm B$, respectively.
We perform a unitary transformation $U(\bm B)$ of the Hamiltonian, so that its eigenstates are aligned with an instantaneous $z$ direction, fulfilling $\sigma^z |\pm\rangle = \pm |\pm\rangle$: 
\begin{equation}
\overline{\cal H}_{S}(\bm B)=U(\bm B){\cal H}_{S}(\bm B)U^\dagger(\bm B)= B\sigma^z.
\end{equation}
The frozen state of the system can be expressed in the Bloch vector representation as
\begin{equation}
    \rho^{(f)}(t)= \frac{1}{2} \left(\sigma^0 + \bm r^{(f)}(t) \cdot \bm \sigma \right),
\end{equation}
being $\sigma^0$ the $2\times 2$ identity matrix. 
For this system, there are only two possible quantum of energies $\pm 2B$ and, consequently, 
the  jump operators~\eqref{L-ops} that appears in the Lindblad equation are
\begin{equation}
L_{\alpha, 2B} = g_\alpha \sqrt{2B \left(n_\alpha(2B)+1\right)}\xi^*_{\alpha}\sigma^-,\quad\quad 
L_{\alpha, -2B} = g_\alpha \sqrt{2B n_\alpha(2B)}\xi_{\alpha}\sigma^+,
\end{equation}
where the ladder operators are $\sigma^{\pm} = (\sigma^x \pm i\sigma^y)/2$ and $\xi_\alpha = \langle +|\pi_\alpha|-\rangle$. 

Using the algebra of Pauli matrices and substituting in the expressions of the dissipators, we get the two terms of the master equation, the unitary evolution
\begin{equation}
-i \left[\overline{\cal H}_{S}(\bm B),\rho^{(f)}(t)\right] = B \left({r}^{(f)}_y(t) 
\sigma^x - {r}^{(f)}_x(t) \sigma^y \right),
\end{equation}
and the dissipative one
\begin{eqnarray}
{\cal D}_{\alpha}\left[\rho^{(f)}(t)\right]  =  
-\Gamma_\alpha(\bm B)  \bigg\{2 \sigma^z + \big(1+2n_\alpha(2B)\big)
\left[r^{(f)}_x(t)\sigma^x+r^{(f)}_y(t)\sigma^y+2r^{(f)}_z(t)\sigma^z\right]\bigg\},
\end{eqnarray}
where $\Gamma_\alpha(\bm B) =  B g_\alpha^2  |\xi_\alpha|^2/2$.

The solution of Eq. (\ref{lind}) in the stationary regime, 
that is 
$\rho^{(f)}(t)\rightarrow e^{i \overline{\cal{H}}_St}\rho^{(f)}(t)e^{-i \overline{\cal{H}}_St} $ and $d\rho^{(f)}/dt=0$, can be expressed as follows,
\begin{equation}\label{ma-bloch}
{\cal M}(\bm B) \; \overline{\boldsymbol{r}}^{(f)}(\boldsymbol{B})=\boldsymbol{\gamma}(\boldsymbol{B}).
\end{equation}
with 
\begin{eqnarray}
{\cal M} (\boldsymbol{B}) &=& \mbox{Diag}\left(
\Gamma^+(\boldsymbol{B}), \Gamma^+(\boldsymbol{B}), 2 \Gamma^+(\boldsymbol{B})\right), \nonumber \\
\boldsymbol{\gamma}(\boldsymbol{B})&=& \left(0,0,2 \Gamma^-(\boldsymbol{B})\right),
\end{eqnarray}
being 
\begin{equation}
\Gamma^{\pm}(\boldsymbol{B})=\sum_{\alpha} \Gamma_{\alpha}(\boldsymbol{B}) \left[
 n_{\alpha}(2B) \pm \left(1+n_{\alpha}(2B)\right) \right].
\end{equation}
Assuming reservoirs at equal temperature, the solution is  
\begin{equation}
\overline{\boldsymbol{r}}^{(f)}(\boldsymbol{B})=\frac{\Gamma^-(\boldsymbol{B})}{\Gamma^+(\boldsymbol{B})}\boldsymbol{n}_z= \tanh(\beta B)\boldsymbol{n}_z,\;\;\;\;\;\;\boldsymbol{n}_z=(0,0,1).
\end{equation}
Hence, we verify that the frozen state is the usual thermal state:
\begin{equation}\label{froz}
\rho^{(f)}(\boldsymbol{B})=\frac{1}{2}\left(\sigma^0+\tanh{\beta B}\sigma^z\right)=\frac{e^{-\beta \overline{\cal H}_{S}(\boldsymbol{B})}}{Z^{(f)}}.
\end{equation}
It is important to notice that this operator is expressed in the instantaneous basis that diagonalizes 
the frozen Hamiltonian for the system. If we express this operator in some other basis (for instance, in a particular laboratory frame), we must implement the transformation
\begin{equation}\label{froz-lab}
\rho^{(f,{\rm lab})}(\boldsymbol{B})=U^\dagger(\boldsymbol{B})\rho^{(f)}(\boldsymbol{B}) U(\boldsymbol{B})=
%\frac{1}{2}U^\dagger(\boldsymbol{B})\left(\sigma^0+\tanh(\beta B)\sigma^z\right)U(\boldsymbol{B})=
\frac{1}{2} \left[ \sigma^0+\tanh(\beta B)\boldsymbol{n}_{\boldsymbol{B}}\cdot \boldsymbol{\sigma}\right],
\end{equation}
where we can identify the Bloch vector for this state in the laboratory frame,
\begin{equation}
    \boldsymbol{r}^{(f,\rm{lab})}=R(\boldsymbol{B}) \overline{r}^{(f)}(\boldsymbol{B})= \tanh(\beta B)\boldsymbol{n}_{\boldsymbol{B}}.
\end{equation}

The adiabatic correction in the instantaneous basis is calculated taking into account the fact that this basis changes as a function of $\boldsymbol{B}$. Hence,
\begin{eqnarray}\label{adia-1qubit}
    {\boldsymbol{r}}^{(1)}(\boldsymbol{B}) &=&{\cal M}^{-1} (\boldsymbol{B})R^\dagger(\boldsymbol{B})\partial_{\boldsymbol{B}}{\boldsymbol{r}}^{(f,\rm{lab})}(\boldsymbol{B}) \cdot 
    \dot{\boldsymbol{B}} = \nonumber \\
    &=&-\frac{\tanh(\beta B)}{2\sum_{\alpha} \Gamma_{\alpha}(\boldsymbol{B})} R^\dagger(\boldsymbol{B})\left[\partial_{\boldsymbol{B}}\left[\tanh(\beta B) \right]\cdot 
    \dot{\boldsymbol{B}}\;\boldsymbol{n}_{\boldsymbol{B}}+ 2 \tanh(\beta B) \partial_{\boldsymbol{B}} \boldsymbol{n}_{\boldsymbol{B}}\cdot 
    \dot{\boldsymbol{B}}\right],
%    {\boldsymbol{r}}^{1,\rm{(lab)}}(\boldsymbol{B}) &=&R(\boldsymbol{B}){\cal M}^{-1} (\boldsymbol{B})R^\dagger(\boldsymbol{B}) \partial_{\boldsymbol{B}}\boldsymbol{r}^{f,\rm{(lab)}}(\boldsymbol{B}) \cdot \dot{\boldsymbol{B}}
    %=-\frac{1}{2\Gamma^+(\boldsymbol{B})} \partial_{\boldsymbol{B}}{\boldsymbol{r}}^{f,(\rm{lab})}(\boldsymbol{B}) \cdot  \dot{\boldsymbol{B}}
\end{eqnarray}
where we notice that 
$\partial_{\boldsymbol{B}} \boldsymbol{n}_{\boldsymbol{B}} \perp \boldsymbol{n}_z$.

Therefore, the first-order correction in the slow-driving expansion is
\begin{equation}
%\label{adia}
\rho^{(1)}(\boldsymbol{B})= {\boldsymbol{r}}^{(1)}(\boldsymbol{B}) \cdot \boldsymbol{\sigma} .
%-\frac{\tanh(\beta B)}{2\sum_{\alpha} \xi_{\alpha}(\boldsymbol{B})} \partial_{\boldsymbol{B}}\left[\tanh(\beta B) \right]\cdot   \dot{\boldsymbol{B}}\; \sigma^z.
\end{equation}

We can now calculate ${\cal D}_\alpha[\rho^{(1)}(\boldsymbol{B})]$
%\begin{equation}
%{\cal D}_\alpha[\rho^{(1)}(\boldsymbol{B})] = -\Gamma_{\alpha}(\boldsymbol{B}) \frac{\tanh(\beta B)}{\sum_{\alpha} \xi_{\alpha}(\boldsymbol{B})}\partial_{\boldsymbol{B}}\tanh(\beta B) \cdot 
%    \dot{\boldsymbol{B}} \;\sigma^z,
%\end{equation}
and, substituting in the expression of the heat current we get
\begin{equation} \label{curr-adia-qubit}
J_\alpha^{(1)}(t) =
%-\frac{  \Gamma_\alpha (\boldsymbol{B})}{\tanh(\beta B)\sum_\alpha \Gamma_\alpha(\boldsymbol{B})} \tanh(\beta B) \partial_{\boldsymbol{B}}\tanh(\beta B) \cdot \dot{\boldsymbol{B}} = 
- \frac{  \Gamma_\alpha(\boldsymbol{B}) }{\sum_\alpha \Gamma_\alpha(\boldsymbol{B}) }  \lm{B} \; \partial_{\boldsymbol{B}}\tanh(\beta B) \cdot 
    \dot{\boldsymbol{B}},
\end{equation}
where we notice that only the first term of Eq. (\ref{adia-1qubit}) contributes.

\lm{Using Eq. \eqref{sdot} we can calculate}
\begin{equation}\label{ds-qubit}
\frac{dS^{(f)}}{dt}= - \lm{\frac{B}{T}}\partial_{\boldsymbol{B}}\tanh(\beta B) \cdot 
    \dot{\boldsymbol{B}} ,
\end{equation}
hence,
\begin{equation}
   J_\alpha^{(1)}(t) = \frac{  \Gamma_\alpha(\boldsymbol{B})  }{\sum_\alpha \Gamma_\alpha(\boldsymbol{B}) } \lm{T} \frac{dS^{(f)}}{dt} .
\end{equation}
% In the laboratory frame, the first order correction reads
% \begin{equation}\label{froz-lab2}
% \rho^{(1,{\rm lab})}(\boldsymbol{B})=\boldsymbol{r}^{(1,\rm{lab})}(\boldsymbol{B})\cdot \boldsymbol{\sigma},
% \end{equation}

\section{Exact analytical solution of Lindblad equation for 
\texorpdfstring{$N_{\mathrm q}$}{Nq} non-interacting driven qubits}
%\section{Exact analytical solution of Lindblad equation for $N_q$ non-interacting driven qubits}
\label{app:many-qubits}

We start considering the Hamiltonian of $N_q$ non-interacting qubits as,
\begin{equation}\label{H_not_corr}
    \mathcal{H}_S = h_1\otimes \mathbb{I}_2 \otimes \dots \otimes \mathbb{I}_{N_q} + \mathbb{I}_1 \otimes h_2  \otimes \dots \otimes \mathbb{I}_{N_q} + \dots + \mathbb{I}_1 \otimes \mathbb{I}_2 \otimes \dots \otimes h_{N_q} , 
\end{equation}
where $h_j$ is the individual Hamiltonian of each qubit.
The density matrix for this system is given by
\begin{equation}
    \rho^{(f)} = \rho^{(f)}_1 \otimes \rho^{(f)}_2 \dots \otimes\rho^{(f)}_{N_q}.
\end{equation}
Using the Leibniz rule to derive and 
% \begin{equation}\label{leibniz_rule}
%     \frac{d\rho^{(f)}}{dt} = \frac{d \rho_1^{(f)}}{dt}\otimes\rho_2^{(f)}\dots \otimes\rho^{(f)}_{N_q} + \rho_1^{(f)}\otimes\frac{d\rho_2^{(f)}}{dt} \otimes \dots \otimes\rho^{(f)}_{N_q} + \cdots
%     %     + \rho_1^{(f)}\otimes\rho_2^{(f)}\otimes\dots \otimes\frac{d \rho_{N_q}^{(f)}}{dt}
% \end{equation}
%
%On the other hand, we can use 
the master equation for each qubit, $q_j$ $\frac{d \rho_j}{dt} = \mathcal{L}_{q_j}[\rho_j]$, 
we can write
\begin{equation}
\label{leibniz_rule}
    \frac{d\rho^{(f)}}{dt} =  \mathcal{L}_{q_1}[\rho_1^{(f)}]\otimes\rho_2^{(f)}\dots \otimes\rho^{(f)}_{N_q} + \rho_1^{(f)}\otimes\mathcal{L}_{q_2}[\rho_2^{(f)}] \otimes \dots \otimes\rho^{(f)}_{N_q} + \cdots 
    %+ \rho_1^{(f)}\otimes\rho_2^{(f)}\otimes\dots \otimes \mathcal{L}_{q_{N_q}}[\rho_{N_q}^{(f)}] .
\end{equation}
To determine the frozen component of the density matrix we put $\frac{d \rho_j^{(f)}}{dt} = 0$ in \eqref{leibniz_rule} and using Eq.~\eqref{ma-bloch} we get
\begin{align}
    0    &=[{\cal M} (\boldsymbol{B}_1) \; \overline{\boldsymbol{r}}^{(f)}_1(\boldsymbol{B}_1) - \boldsymbol{\gamma}(\boldsymbol{B}_1)]\cdot \boldsymbol{\sigma}\otimes\rho_2^{(f)}\dots \otimes\rho^{(f)}_{N_q}  + \cdots     \notag\\
    &  \:\:\:\:\:\cdots + \rho_1^{(f)}\otimes\dots \otimes [{\cal M} (\boldsymbol{B}_{N_q}) \; \overline{\boldsymbol{r}}^{(f)}_{N_q}(\boldsymbol{B}_{N_q}) - \boldsymbol{\gamma}(\boldsymbol{B}_{N_q})]\cdot \boldsymbol{\sigma} .
\end{align}
Since the frozen density operators are non-zero, it follows that,
\begin{equation}
    {\cal M} (\boldsymbol{B}_j) \; \overline{\boldsymbol{r}}^{(f)}_j(\boldsymbol{B}_1) - \boldsymbol{\gamma}(\boldsymbol{B}_j)=0 \:\:\:\: j=1,2 \dots N_q.
\end{equation}
 From this, we can calculate the frozen Bloch vector for each qubit, 
% \begin{equation}
%     \boldsymbol{r}^{(f)}_j(\bm B_j) = {\cal M} (\bm B_j)^{-1} \; \boldsymbol{\gamma}(\bm B_j) \:\:\:\:j=1,2 \dots N_q,
% \end{equation}
and express $\rho^{(f)}_j$ as
\begin{equation}\label{rhoF_j}
    \rho^{(f)}_j = \frac{\sigma_0}{2} + \frac{1}{2} {\cal M} (\bm B_j)^{-1} \; \boldsymbol{\gamma}(\bm B_j) \cdot \boldsymbol{\sigma} \:\:\:\:j=1,2 \dots N_q.
\end{equation}
Replacing the expressions in appendix \ref{app:single-qubit} we get,
\begin{equation}\label{froz2}
\rho^{(f)}_j(\bm B_j)=\frac{1}{2}\left(\sigma^0+\tanh{\beta B_j}\sigma^z\right)=\frac{e^{-\beta h_j(\boldsymbol{B_j})}}{Z} \:\:\:\:j=1,2,\dots,N_q.
\end{equation}
Now we want to determine $\rho^{(1)}$; to this end, we consider the first-order terms in the Lindblad equation. We replace our result \eqref{froz2} for $\rho^{(f)}_j$ in Eq.~\eqref{leibniz_rule} and solve
\begin{equation}\label{ec_order_1}
    \frac{d\rho^{(f)}}{dt} =  \mathcal{L}_{q_1}[\rho_1^{(1)}]\otimes\rho_2^{(f)}\dots \otimes\rho^{(f)}_{N_q} +  \cdots + \rho_1^{(f)}\otimes\rho_2^{(f)}\otimes\dots \otimes \mathcal{L}_{q_{N_q}}[\rho_{N_q}^{(1)}] .
\end{equation}
%\begin{equation}
%    \frac{d \rho_1^{(f)}}{dt}\otimes\rho_2^{(f)}\dots \otimes\rho^{(f)}_{N_q} + \cdots + \rho_1^{(f)}\otimes\rho_2^{(f)}\otimes\dots \otimes\frac{d \rho_{n}^{(f)}}{dt} =  \mathcal{L}_{q1}[\rho_1^{(1)}]\otimes\rho_2^{(f)}\dots \otimes\rho^{(f)}_{N_q} + \cdots + \rho_1^{(f)}\otimes\rho_2^{(f)}\otimes\dots \otimes \mathcal{L}_{q_{n}}[\rho_{n}^{(1)}],
%\end{equation}
This is equivalent to $N_q$ equations analogous to \eqref{adia-1qubit} for each qubit.

To finalize, we calculate the first order heat current. From the right hand side of Eq.\eqref{ec_order_1} we know
\begin{equation}\label{Diss_2qb_no_corr}
    \mathcal{D}_\alpha[\rho^{(1)}] = \mathcal{D}_{q_1}[\rho_1^{(1)}]\otimes\rho_2^{(f)} \otimes \dots \otimes \rho_{N_q}^{(f)} + \dots +\rho_1^{(f)} \otimes \dots \otimes \mathcal{D}_{q_{N_q}}[\rho_{N_q}^{(1)}].
\end{equation}
Thus, replacing Eqs. \eqref{H_not_corr} and \eqref{Diss_2qb_no_corr} in the first-order heat current
$J_\alpha = \mathrm{Tr}\{ \mathcal{D}_\alpha[\rho^{(1)}] \mathcal{H}_S \}$
we find,
%\begin{align}
%    J_\alpha &= \mathrm{Tr}\{ \mathcal{D}_\alpha[\rho^{(1)}] \cdot \mathcal{H}_S \} ,\\
%    &= \mathrm{Tr}_{1}\dots\mathrm{Tr}_{n}\Big\{ \left(\mathcal{D}_{q_1,\alpha}[\rho^{(1)}_1] \otimes \dots \otimes \rho^{(f)}_n +\dots + \rho^{(f)}_1 \otimes \dots \otimes \mathcal{D}_{q_n,\alpha}[\rho^{(1)}_n] \right)  (h_1\otimes \dots \otimes \mathbb{I}_n + \dots + \mathbb{I}_1 \otimes\dots\otimes h_n)  \Big\} ,\\
%    &= \mathrm{Tr}_{1}\dots\mathrm{Tr}_{n}\left\{ \mathcal{D}_{q_1,\alpha}[\rho^{(1)}_1] \; h_1 \otimes \dots \otimes \rho_n^{(f)} + \dots + \rho^{(f)}_1 \otimes \dots \otimes \mathcal{D}_{q_n,\alpha}[\rho^{(1)}_n] \; h_n   \right\} ,\\
%    &= \mathrm{Tr}_{1} \left\{ \mathcal{D}_{q_1,\alpha}[\rho^{(1)}_1] \; h_1 \right\} \mathrm{Tr}_{2}\{ \rho^{(f)}_2\} \dots \mathrm{Tr}_{n}\{ \rho^{(f)}_n \} + \dots + \mathrm{Tr}_{1}\{\rho^{(f)}_1  \} \mathrm{Tr}_{2}\{ \rho^{(f)}_2\} \dots \mathrm{Tr}_{n}\left\{ \mathcal{D}_{q_n,\alpha}[\rho^{(1)}_n] \; h_n   \right\} ,\\
%    &= J_{q_1,\alpha}^{(1)} + \dots + J_{q_n,\alpha}^{(1)} \leq k_B T \left( \frac{dS^{(f)}_{q_1}}{dt} + \dots + \frac{dS^{(f)}_{q_n}}{dt} \right),
%\end{align}
\begin{eqnarray}
 J_\alpha & = & \mathrm{Tr}_{1}\dots\mathrm{Tr}_{N_q}\Big\{ \left(\mathcal{D}_{q_1,\alpha}[\rho^{(1)}_1] \otimes \dots \otimes \rho^{(f)}_n +\dots + \rho^{(f)}_1 \otimes \dots \otimes \mathcal{D}_{q_{N_q},\alpha}[\rho^{(1)}_{N_q}] \right) 
 {\cal H}_{\rm S} \Big\} = \notag\\
    % \notag \\
    % &\times(h_1\otimes \dots \otimes \mathbb{I}_n + \dots + \mathbb{I}_1 \otimes\dots\otimes h_{N_q})  
    % & = & \mathrm{Tr}_{1}\dots\mathrm{Tr}_{N_q}\left\{ \mathcal{D}_{q_1,\alpha}[\rho^{(1)}_1] \; h_1 \otimes \dots \otimes \rho_{N_q}^{(f)} + \dots + \rho^{(f)}_1 \otimes \dots \otimes \mathcal{D}_{q_{N_q},\alpha}[\rho^{(1)}_{N_q}] \; h_{N_q}   \right\} , \notag\\
     & = & \mathrm{Tr}_{1} \left\{ \mathcal{D}_{q_1,\alpha}[\rho^{(1)}_1] \; h_1 \right\} \dots \mathrm{Tr}_{N_q}\{ \rho^{(f)}_{N_q} \} + \dots + \mathrm{Tr}_{1}\{\rho^{(f)}_1  \} \dots \mathrm{Tr}_{N_q}\left\{ \mathcal{D}_{q_{N_q},\alpha}[\rho^{(1)}_{N_q}] \; h_{N_q}   \right\}.\nonumber
     \end{eqnarray}
     Consequently, 
     \begin{equation}
     J_{q_1,\alpha}^{(1)} + \dots + J_{q_{N_q},\alpha}^{(1)} \leq \lm{T} \left( \frac{dS^{(f)}_{q_1}}{dt} + \dots + \frac{dS^{(f)}_{q_{N_q}}}{dt} \right).
\end{equation}
The heat currents may be different for each qubit, but each of them is bounded as derived for Eq.~\eqref{j-s-nonint}. Thus, when integrating over some trajectory in the parameter space, the heat pump is bounded by
\begin{equation}
    Q_{N_q\;\mathrm{qubits}} = \sum_{j=1}^{N_q} Q^{(\mathrm{pump})}_j \leq  \sum_{j=1}^{N_q} \lm{T} \Delta S^{(f)}_j \leq N_q \; k_B T \ln 2.
\end{equation}
\section{Heat current for two interacting qubits} \label{ent}

We write a general interacting qubits Hamiltonian as follows,
\begin{equation}
    \mathcal{H}_S = h_1\otimes \sigma_0 + \sigma_0 \otimes h_2 + h_{\rm int} .
\end{equation}
The density matrix of two correlated qubits does not factorize into a tensor product of 
the individual single-qubit density matrices. By associating to each qubit a Bloch 
vector $\bm r^{(f)}_{q_i}$, $i=1,2$, the correlated contribution can be separated 
explicitly, allowing us to write
\begin{equation}
    \rho^{(f)} = \frac{1}{4} \left( \mathbb{I}_4 + \sum_{i=1}^3 r_{q_1,i}^{(f)} \; \sigma_{i} \otimes \sigma_{0} + \sum_{i=1}^3 r_{q_2,i}^{(f)} \; \sigma_{0} \otimes \sigma_{i} + \sum_{i,j=1}^3 (r_{q_1,i}^{(f)} r_{q_2,j}^{(f)} + \Delta R_{ij}^{(f)}) \; \sigma_{i} \otimes \sigma_{j} \right),
\end{equation}
where we define the correlated contribution to $\rho^{(f)}$
\begin{equation}
    \rho_{12}^{(f)} = \sum_{i,j=1}^3 \Delta R_{ij}^{(f)} \, \sigma_{i} \otimes \sigma_{j},
\end{equation}
so that
\begin{equation}\label{correlated_qub_mat}
    \rho^{(f)} = \rho^{(f)}_{q_1} \otimes \rho^{(f)}_{q_2} + \rho^{(f)}_{12}.
\end{equation}

To know how the dissipator acts in $\rho^{(1)}$ we take the time derivative of Eq.~\eqref{correlated_qub_mat} and use the Leibniz rule,
\begin{equation}
    \frac{d\rho^{(f)}}{dt} = \frac{d \rho^{(f)}_{q_1}}{dt} \otimes \rho^{(f)}_{q_2} + \rho^{(f)}_{q_1} \otimes \frac{d \rho^{(f)}_{q_2}}{dt} + \frac{d\rho^{(f)}_{12}}{dt}.
\end{equation}
Thus, replacing the Lindblad equation in each time derivative we get
\begin{align}
    \frac{d\rho^{(f)}}{dt} = \mathcal{L}_{q_1}[\rho_{q_1}^{(1)}] \otimes \rho^{(f)}_{q_2} + \rho^{(f)}_{q_1} \otimes \mathcal{L}_{q_2}[\rho_{q_2}^{(1)}] + \mathcal{L}[\rho^{(f)}_{12}] .
\end{align}
Then, using that $\mathcal{L} = \sum_{\alpha} \mathcal{D}_\alpha$ we identify the action of the dissipator over $\rho^{(1)}$,
\begin{equation}
    \mathcal{D}_\alpha[\rho^{(1)}] = \mathcal{D}_{q_1,\alpha}[\rho_{q_1}^{(1)}] \otimes \rho_{q_2}^{(f)} + \rho_{q_1}^{(f)} \otimes \mathcal{D}_{q_2,\alpha}[\rho_{q_1}^{(1)}] + \mathcal{D}_{\alpha}[ \rho^{(1)}_{12} ].
\end{equation}
Finally, replacing this results in the heat current we find
% \begin{equation}
%     J_\alpha =  \mathrm{Tr}\bigg\{ \left( \mathcal{D}_{q_1,\alpha}[\rho_{q_1}^{(1)}] \otimes \rho_{q_2}^{(f)} + \rho_{q_1}^{(f)} \otimes \mathcal{D}_{q_2,\alpha}[\rho_{q_2}^{(1)}] + \mathcal{D}_{\alpha}[ \rho^{(1)}_{12} ] \right)  (h_1\otimes \sigma_0 + \sigma_0 \otimes h_2 + h_{int} ) \bigg\},
% \end{equation}
%
\begin{multline}\label{current_interac_qubits}
    J^{(1)}_\alpha = \mathrm{Tr}_1\left\{ \mathcal{D}_{q_1,\alpha}[\rho_{q_1}^{(1)}] \; h_1 \right\} + \mathrm{Tr}_2\left\{ \mathcal{D}_{q_2,\alpha}[\rho_{q_2}^{(1)}] \; h_2 \right\}+ \\
    + \mathrm{Tr}\left\{ \mathcal{D}_{q_1,\alpha}[\rho_{q_1}^{(1)}] \otimes \rho_{q_2}^{(f)} \:h_{\rm int} \right\} 
    + \mathrm{Tr}\left\{ \rho_{q_1}^{(f)} \otimes \mathcal{D}_{q_2,\alpha}[\rho_{q_2}^{(1)}] \:  h_{int} \right\} 
    + \mathrm{Tr}\left\{ \mathcal{D}_{\alpha}[ \rho^{(1)}_{12} ] \: \mathcal{H}_S \right\} .
\end{multline}
We identify the last three terms in \eqref{current_interac_qubits} with the correlation between the two qubits, thus we denote them by $J^{(1)}_{\mathrm{12},\alpha}$ and write,
\begin{equation}\label{current_2qb_interacting}
     J^{(1)}_\alpha = J^{(1)}_{q_1,\alpha} + J^{(1)}_{q_2,\alpha} + J^{(1)}_{\mathrm{12},\alpha} .
\end{equation}
The last term in Eq.~\eqref{current_2qb_interacting} does not have a defined sign, 
so, for a single reservoir, the heat current may exceed $T ( d S_{1}^{(f)}/dt + d S^{(f)}_2/dt)$, 
showing that the Landauer bound does not apply for the heat current associated with 
an individual reservoir. Nevertheless,  Eq.~\eqref{demo_balance1} guarantees that the Landauer bound is 
fulfilled by the total contribution $\sum_\alpha J_\alpha^{(1)}$.

\end{appendix}

\bibliography{SciPost_Example_BiBTeX_File.bib}

\end{document}